\documentclass[aps,prx,twocolumn,superscriptaddress]{revtex4-2}
\usepackage{amsmath,bbm,amssymb,amsfonts,bm,color,graphicx,tabularx,physics}
\usepackage{balance}
\usepackage[unicode=true,colorlinks=true]{hyperref}
\usepackage[etex=true,export]{adjustbox}

\hypersetup{linkcolor=blue,citecolor=blue,urlcolor=blue}
\usepackage{diagbox}
\usepackage{color}

\newcommand{\beq}{\begin{equation}}
	\newcommand{\eeq}{\end{equation}}
\newcommand{\beql}{\begin{equation*}}
	\newcommand{\eeql}{\end{equation*}}
\newcommand{\beqn}{\begin{eqnarray}}
	\newcommand{\eeqn}{\end{eqnarray}}

\renewcommand{\vec}[1]{\mbox{\boldmath$#1$}}
\usepackage{comment}
\usepackage{booktabs}
\usepackage{mathrsfs}

\begin{document}
	\title{Symmetry-Based Real-Space Framework for Realizing Flat Bands and Unveiling Nodal-Line Touchings}
	
	\author{Rui-Heng Liu}
	\email{physrhl@gmail.com}
	\affiliation{School of Physics and Wuhan National High Magnetic Field Center, Huazhong University of Science and Technology, Wuhan 430074, China}
	
	\author{Xin Liu}
	\email{phyliuxin@hust.edu.cn}
	\affiliation{School of Physics and Wuhan National High Magnetic Field Center, Huazhong University of Science and Technology, Wuhan 430074, China}
	\affiliation{Tsung-Dao Lee Institute and School of Physics and Astronomy, Shanghai Jiao Tong University, Shanghai 201210, China}
	\affiliation{Institute for Quantum Science and Engineering and Hubei Key Laboratory of Gravitation and Quantum Physics, Huazhong University of Science and Technology, Wuhan 430074, China}
	\affiliation{Hefei National Laboratory, Hefei 230088, China}
	\affiliation{Wuhan Institute of Quantum Technology, Wuhan 430206, China}
	
	\date{\today}
	
	\begin{abstract}
		Flat band (FB) systems provide ideal playgrounds for studying correlated physics, whereas multi-orbital characteristics in real materials are distinguished from most simple FB models. Here, we propose a systematic and versatile framework for FB constructions in tight-binding (TB) models based on symmetric compact localized states (CLSs), integrating lattice and orbital degrees of freedom. In our framework, we first demonstrate that any CLS can be symmetrized into a representation of the point group, which remains valid for high orbitals with finite spin-orbit couplings (SOCs). Second, we determine the candidate CLS sites according to lattice symmetry, and simplify the hopping as a linear mapping between two Hilbert spaces: one of CLS sites and another of their adjacent sites. The existence of FBs depends on a non-empty kernel of the mapping. Finally, we distinguish eigenstates in the kernel to qualify as a CLS. Since both CLSs and Hamiltonian respect symmetries, group theory naturally manifests as a powerful tool to obtain CLSs throughout the procedure, free from the constraints of specific lattices. To illustrate the versatility of our framework, we construct three representative FB models: one in two-dimensional (2D) and the rest in three-dimensional (3D). All of them lack special lattice structures and incorporate high orbitals. Notably, we unveil that 3D FBs can exhibit not only band touchings at points but also along lines, a feature of significant physical interest. For a comprehensive understanding, we derive a concise criterion for determining band touchings, which demonstrates their symmetry-protected nature and provides natural explanations for the occurrence of gapped or gapless FBs. By unifying symmetry principles in real space, our work offers a systematic approach to constructing FBs across diverse lattice systems. This framework opens new avenues for understanding and engineering FB systems, with potential implications for correlated quantum phenomena and exotic phases of matter.
	\end{abstract}
	
	\maketitle
	
	\section{Introduction}\label{sec_intro}
	Flat bands (FBs) in electronic systems have become a fundamental topic in condensed matter physics, where the nearly vanishing bandwidth signifies the quenching of kinetic energy. Therefore, Coulomb interactions prevail and become the dominant energy scale in such systems, leading to exotic quantum phases driven by strong electron-electron correlations \cite{liu2014exotic,checkelsky2024flat}. Early examples include ferromagnetism \cite{mielke1991ferromagnetic,mielke1991ferromagnetism,mielke1992exact,mielke1993ferromagnetism,tasaki1992ferromagnetism,tasaki1998nagaoka,lin2018flatbands,hase2018possibility,bouzerar2023flat,hase2023flat} and (generalized) Wigner crystal \cite{shayegan2022wigner,hubbard1978generalized,wu2007flat,jaworowski2018wigner,regan2020mott,padhi2021generalized}. Recent advances in topology and twistronics have further brought new vitality into the study of FB physics. Twisted systems with tunable band structure and topology manifest as an ideal platform for studying novel quantum states \cite{regan2020mott,padhi2021generalized,tang2020simulation,wang2020correlated,cao2018unconventional,cao2018correlated,balents2020superconductivity,xie2020topology,torma2022superconductivity,andrews2020fractional,park2023observation,cai2023signatures,xu2023observation,lu2024fractional,angeli2021gamma,xian2021realization,claassen2022ultra}, where primitive translation symmetry is broken by Moire potentials, leading to negligible bandwidth in the folded Moire Brillouin zone (BZ). Fascinating unconventional superconductivity \cite{cao2018unconventional,cao2018correlated,balents2020superconductivity,xie2020topology,torma2022superconductivity,iglovikov2014superconducting,peotta2015superfluidity,hofmann2020superconductivity,peri2021fragile} and the fractional quantum Hall effect \cite{andrews2020fractional,park2023observation,cai2023signatures,xu2023observation,lu2024fractional,sun2011nearly,neupert2011fractional,tang2011high,regnault2011fractional} are predicted and confirmed in systems with topological FBs, garnering significant attention in recent years. These novel phenomena motivate the elaborations on constructing strict FBs within a comprehensive framework \cite{mielke1991ferromagnetic,mielke1991ferromagnetism,mielke1992exact,mielke1993ferromagnetism,tasaki1992ferromagnetism,tasaki1998nagaoka,chiu2020fragile,liu2025non,tasaki1992ferromagnetism,tasaki1998nagaoka,lieb1989two,regnault2022catalogue,cualuguaru2022general,maimaiti2017compact,maimaiti2019universal,maimaiti2021flat,mizoguchi2019molecular,mizoguchi2020systematic,mizoguchi2021flat,mizoguchi2023molecular,morales2016simple,rontgen2018compact,hwang2021general,graf2021designing,lee2019hidden,xu2020building,morfonios2021flat,chen2023decoding}, which could provide valuable insights on the single-particle level before further intricate studies on correlation effects. Plethora of schemes have been proposed, many of which utilize special lattice structures, such as the line-graph construction \cite{mielke1991ferromagnetic,mielke1991ferromagnetism,mielke1992exact,chiu2020fragile,chiu2020fragile,liu2025non}, the cell construction \cite{tasaki1992ferromagnetism,tasaki1998nagaoka} and bipartite lattices \cite{lieb1989two,regnault2022catalogue,cualuguaru2022general}. Other proposals include the FB generators \cite{maimaiti2017compact,maimaiti2019universal,maimaiti2021flat}, the molecular-orbital representation \cite{mizoguchi2019molecular,mizoguchi2020systematic,mizoguchi2021flat,mizoguchi2023molecular} and other more \cite{morales2016simple,rontgen2018compact,hwang2021general,graf2021designing,lee2019hidden,xu2020building,morfonios2021flat,chen2023decoding,neves2024crystal,li2025general}. However, most of these methods focus on two-dimensional (2D) lattices while the majority of realistic materials are three-dimensional (3D). In 2D systems, the relation between strict FBs and topology has been well understood. Band touchings play a significant role \cite{bergman2008band,rhim2019classification}
	and introducing proper perturbations to lift the degeneracy at singular touching points is regarded as one of the paradigms to achieve topological FBs \cite{rhim2019classification,herzog2024topological}. It has also been proven that a strict gapped FB exhibits a vanishing Chern number \cite{chen2014impossibility}, yet it is closely related to fragile topology \cite{cualuguaru2022general}. In contrast, such correspondence in 3D systems is more vague. Achieving 3D FBs has ignited interest in recent studies \cite{jiang2021giant,wakefield2023three,hase2024new}, but remains more challenging and elusive both in theory and experiment. 
	
	The formation of FBs is attributed to destructive interference of wavefunctions \cite{bergman2008band,liu2014exotic} and the localization of electrons \cite{sutherland1986localization}, such as in the Kagome \cite{mielke1991ferromagnetic,mielke1991ferromagnetism,mielke1992exact} and Lieb \cite{lieb1989two} lattices with isotropic hopping from $s$-orbitals, as well-known examples. Notably, the multi-orbital characteristics are ubiquitous in realistic quantum materials and most strongly correlated systems are characterized by high-orbital electrons with finite spin-orbit couplings (SOCs). Therefore, many FB models with specific lattice geometries can not be directly applied to realistic materials. Interestingly, high-orbital characteristics also emerge in Moire bands of twisted transition metal dichalcogenide bilayers \cite{angeli2021gamma,xian2021realization,claassen2022ultra}, which is reminiscent of the earlier proposed $p$-orbital honeycomb model in optical lattices \cite{wu2007flat,wu2008p}. The inherent anisotropy of high-orbitals and the interplay between the lattice and orbital degrees of freedom could provide new possibilities for exploring FBs. It has been demonstrated that the existence of corresponding compact localized states (CLSs) is universal to a strict dispersionless band over the whole BZ \cite{bergman2008band,read2017compactly,rhim2019classification}. While some FB construction schemes based on the CLS are proposed in recent studies \cite{maimaiti2017compact,maimaiti2019universal,maimaiti2021flat,morales2016simple,rontgen2018compact,hwang2021general,graf2021designing}, a general and intuitive real-space method incorporating high-orbital CLSs remains lacking. 
	
\begin{figure*} [htbp]
		\centering
		\includegraphics[width= 2\columnwidth]{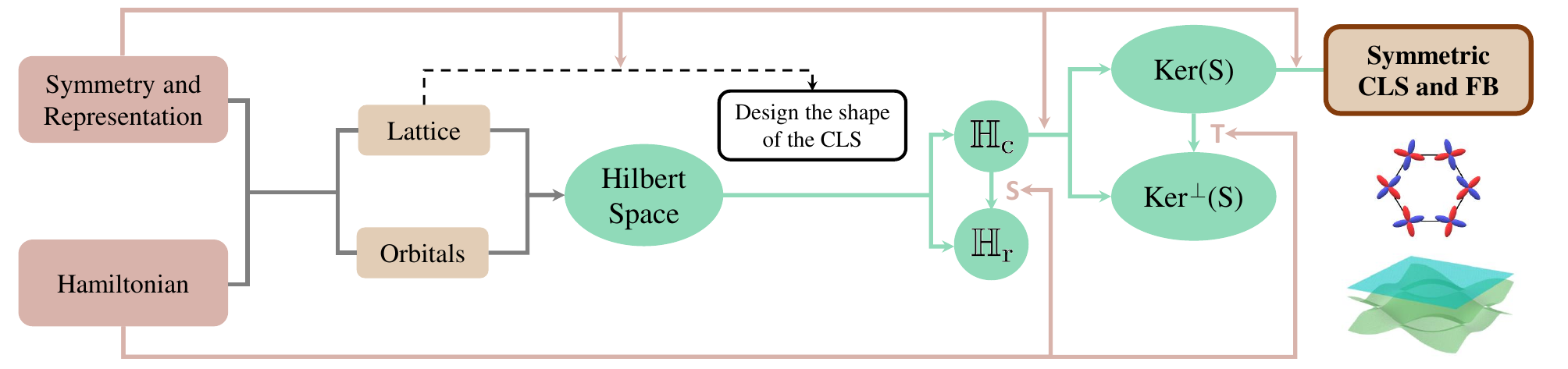}
		\caption{Real space recipe for constructing FBs in this work. It applies to general lattice structures and atomic orbitals. The key is to identify a symmetric CLS.}
		\label{intro}
	\end{figure*}

	In this work, we propose a systematic framework for FB construction that focuses on real space. We utilize group theory and corresponding representations to establish a complete procedure (Fig.~\ref{intro}) to obtain symmetric CLSs in generic tight-binding (TB) models, which is applicable to systems in all dimensions and with different lattice symmetries. It could incorporate practical high orbitals and SOCs as well. With orbital degree of freedom, we demonstrate that FB models are not limited to lattices with special geometry. We employ the Slater-Koster (SK) formalism \cite{slater1954simplified} as a physical and reasonable approximation, and exemplify our scheme with a 2D honeycomb model with $d$-orbitals and a 3D simple cubic model with $s,p$-orbitals. We also employ 2D CLSs to study FBs in Van der Waals stacking structures, providing a new perspective. In addition, we thoroughly investigate the band touchings of FBs by generalizing the concept of ``real space topology" \cite{bergman2008band} with the aid of symmetries. We show that the band touchings are related to the structure group (defined in section \ref{sec_touching}) of the CLS, which is completely determined by the real-space wavefunction. We tabulate the structure groups for all the possible symmetric CLSs and further set up a concise criterion to determine the touchings utilizing representation theory, which also accounts for intriguing  nodal-line touchings in our 3D simple cubic model. 
	
	\section{General framework}\label{sec_framework}
	The most prominent feature of a strict FB lies in its macroscopically‌ degenerate subspace in the momentum space, where the degeneracy matches the number of states within the BZ. Such a high degree of degeneracy allows us to construct specific real-space eigenstates by an appropriate superposition of all the degenerate Bloch states $\psi_{\boldsymbol{k}}$. These eigenstates are known as the CLS, which is a special type of Wannier-like function confined to a finite number of lattice sites and completely vanishing outside this region \cite{read2017compactly,bergman2008band,rhim2019classification}. Notably, the FBs discussed here are distinct from those in the atomic limit, whose vanishing bandwidth results from the absence of wavefunction overlap between neighboring sites. In contrast, the specific distribution of the CLS leads to destructive interference, thereby preventing it from hopping outwards and rendering it as an exact eigenstate \cite{bergman2008band,liu2014exotic}. As the distinction between FBs in the atomic limit and induced by quantum interference is clearly reflected in real-space properties of the localized states, we derive a general theoretical framework for FB by examining the CLS as a universal characteristic. Our construction framework will naturally apply to any crystal system and orbitals residing on different Wyckoff positions (WPs). Formally, we focus on the following TB Hamiltonian with finite-range hoppings as
	\begin{equation}\label{H}
		\begin{split}
			\hspace{-0.75em}
			H=&
			\sum_{\alpha\beta}\left(\sum_{\langle ij\rangle}t^{\langle ij\rangle}_{\alpha\beta} c^{\dagger}_{i\alpha}c_{j\beta}+
			\sum_{\langle\langle ij\rangle\rangle}t^{\langle\langle ij\rangle\rangle}_{\alpha\beta} c^{\dagger}_{i\alpha}c_{j\beta}+\cdots\right)\\
			&+i\sum_{\alpha\beta}\sum_{ij}\lambda^{ij}_{\alpha\beta}c^{\dagger}_{i\alpha}c_{j\beta}.
		\end{split}
	\end{equation}	
	Here, $i,j$ labels the sites. $t^{\langle ij\rangle}\text{ }(t^{\langle\langle ij\rangle\rangle})$ is the (next) nearest-neighbor hopping matrix with $\alpha,\beta$ labeled internal degree of freedom (orbital, spin, etc.). The last term represents the possible SOC. We first sketch the flow as shown in Fig.~\ref{intro}. Obtaining a symmetric CLS is essential in our construction. To achieve that, we first design the shape of the CLS according to the lattice symmetry, which provides the candidate CLS sites. Then, we treat the Hamiltonian as a mapping $S$ from these sites to their neighbors. The existence of FB relies on a non-empty kernel $\text{Ker}(S)$. Also, following elaborations are required to distinguish eigenstates in the kernel, which qualify them as CLSs. In the whole process, we classify all the states according to the irreducible representation (Irrep) of the symmetry group to facilitate the procedure and finally obtain parameters producing FBs. Remarkably, these Irreps can naturally provide a new classification of all the FB models, which also has further indications on the physical origin of the FBs.
	
	\subsection{Construction of symmetric CLS}
	Identifying a single CLS is sufficient to reproduce the whole Hilbert space of the FB, since we can generate a family of them by translating the Bravais-lattice (BL) vector $\boldsymbol{R}$ (these CLSs may be incomplete in some FB models, see section \ref{sec_touching} for details). Due to the macroscopically‌-degenerate nature of the FB, it is important to notice that a symmetric CLS under point group symmetry always exists. For a set of Wannier-like functions, they are called symmetric if they form a representation of the considered group \cite{kruthoff2017topological,po2017symmetry,bradlyn2017topological,cano2021band,hwang2021flat,schindler2021noncompact}. For the sake of clarity, we denote the space group of the system as $\mathcal{S}$, and $\mathcal{G}$ is a point group and also a subgroup of $\mathcal{S}$. Suppose a FB is characterized by $\Phi_{\text{CLS}}$ that does not respect $\mathcal{G}$, we can generate a set of states as $\varphi_{i}=g_{i}\Phi_{\text{CLS}}$ where $g_{i}$ is the operator corresponds to a point group element. These states are closed under all the group elements and naturally carry a representation. Also since the Hamiltonian has the symmetry, $[H,g_{i}]=0$ so that each $\varphi_{i}$ is still a localized eigenstate with identical energy. From the above analysis, we can conclude that it is always possible to impose proper linear combinations of $\varphi$, resulting in new eigenstates that carry Irreps of the point group. From this viewpoint, they are similar to molecular orbitals with point group symmetry but not translation symmetry. For simplicity, we consider a symmorphic space group, which allows us to obtain $\mathcal{G}$ by $\mathcal{G}=\mathcal{S}/\mathcal{T}$ as a quotient group with $\mathcal{T}$ the normal translation subgroup. Hereafter, we denote the symmetric CLSs as $\ket{w(\boldsymbol{R})}$ with $\boldsymbol{R}$ as the center of the state, and they serve as the backbone of our scheme.
	
	As a symmetric CLS, its shape should be compatible with the symmetry. Specifically, this indicates that the Hilbert space of the occupied sites is closed under all the point group operations. To construct the symmetric CLS, it is necessary to first specify the origin of all the operations. We choose the WPs (of the space group $\mathcal{S}$) with unit multiplicity, e.g. 1a WP, which can realize the full point group symmetry. Subsequently, we design the shape of the CLS by generating a set of lattice sites from a single site adjacent to the origin, denoted as $\vec{q}$, as
	\begin{equation}\label{orb}
		\text{Orb}_{\boldsymbol{q}}=\{O_{g}\vec{q}|g\in\mathcal{G}\}.
	\end{equation}
	Here, $O_{g}$ is the $d$-dimensional spatial transformation of the point group operation. $\vec{q}$ could occupy any WP according to the specific lattice structure. We denote the distinct sites as $\{\vec{\delta}_{i}\}$, and all these sites share a conjugated stabilizer group and hence belong to the same WP \cite{cano2021band}. As a further physical requirement, we
	demand that $\vec{q}$ should not coincide with the origin, otherwise $\{\vec{\delta}_{i}\}$ only contains one site and this is exactly the atomic limit.  This procedure naturally provides the minimal candidate of the symmetric CLS, and we denote the corresponding Hilbert space as $\mathbb{H}_{\text{c}}$. Formally, the set (Eq.~\eqref{orb}) is referred to as a $\mathcal{G}$-orbit \cite{supp}, and in general, it is allowed to design the shape of the CLS as the union of several $\mathcal{G}$-orbits. However, it is always convenient to start from the simplest one. After specifying the candidate CLS sites, the total Hilbert space of the system is partitioned as $\mathbb{H}_{\text{c}}\oplus\mathbb{H}_{\text{r}}$, i.e., a direct sum of the subspace spanned by the orbitals on candidate CLS sites and the rest of the system. Hence, the overall Hamiltonian takes
	\begin{equation}\label{S}
		H=
		\begin{bmatrix}
			H_{\text{c}} & S^{\dagger} \\
			S & H_{\text{r}}
		\end{bmatrix}
	\end{equation}
	with $S$ the hopping from $\mathbb{H}_{\text{c}}$ to $\mathbb{H}_{\text{r}}$. The wavefunction of the symmetric CLS should take $\ket{w(\boldsymbol{R})}=\{\psi,0\}$ in the basis $\{\mathbb{H}_{\text{c}},\mathbb{H}_{\text{r}}\}$. Accordingly, we set up a general criterion for a CLS as
	\begin{equation}\label{cond}
		S\psi=0 \qquad H_{\text{c}}\psi=E_{\text{FB}}\psi
	\end{equation}
	which are equivalent to the Schrodinger equation $H\ket{w(\boldsymbol{R})}=E_{\text{FB}}\ket{w(\boldsymbol{R})}$. The FB in the atomic limit is a special case that satisfies these two conditions, but we are interested in the case where $S$ has non-zero hopping elements. In realistic TB models, the hoppings should vanish when two orbitals are far enough, therefore $\mathbb{H}_{\text{r}}$ is truncated from a macroscopic number of lattice sites to a few ones adjacent to $\mathbb{H}_{\text{c}}$, which becomes much smaller and denoted as $\mathbb{H}_{\text{tr}}$. From the above analysis, $S$ could be simplified as a linear mapping between these two Hilbert subspaces of only several sites. The first condition in Eq.~\eqref{cond} mathematically shows that the CLS is in the kernel of $S$. It also has a clear physical picture, which indicates that the CLS associated with a FB is trapped in the region as designed and exhibits a vanishing hopping amplitude outward. We refer to it as the interference condition. Notice that with longer-range hoppings included, the dimension of $\mathbb{H}_{\text{tr}}$ grows rapidly, resulting in a significant increase in the complexity of $S$. Consequently, constructing a model with a non-empty $\text{Ker}(S)$ becomes exceedingly challenging. This complexity elucidates why practical materials can only support narrow bands instead of strict FBs. For the sake of theoretical simplicity, we will mainly focus on models with only nearest-neighbor hoppings $t^{\langle ij\rangle}$ in later examples. In practice, the general procedure to produce a  symmetric CLS can be summarized into two main steps: first, we need to find a proper parameter regime that furnishes a non-empty kernel to allow the CLS to live within. All the states in $\text{Ker}(S)$ are decoupled from $\mathbb{H}_{\text{r}}$ and serve as the candidates of the symmetric CLS; second, we formally have $\mathbb{H}_{\text{c}}=\text{Ker}(S)\oplus\text{Ker}^{\perp}(S)$ with $\text{Ker}^{\perp}(S)$ the orthogonal complement of the kernel, and it is necessary to decouple the candidates from $\text{Ker}^{\perp}(S)$ to get eigenstates, finally resulting in CLSs and FBs.
	
	\subsection{Classification of FBs and physical indication}
	As mentioned earlier, symmetry will play a key role in the above two steps. Our setup naturally guarantees that both $\mathbb{H}_{\text{c}}$ and $\mathbb{H}_{\text{tr}}$ are closed under symmetry operations. Therefore, we can implement both steps with the aid of symmetry representations. In the first step, we can decompose $\mathbb{H}_{\text{c}}$ and $\mathbb{H}_{\text{tr}}$ based on the Irreps of $\mathcal{G}$ (see Eq.~\eqref{decomposition0} as the first example). Notice that only the couplings between the same Irreps are allowed by symmetry, the mapping $S$ is classified into different channels labeled by the Irrep. Since such ``selection rule" is not limited to special lattices, our approach also goes beyond the limitation of lattice geometry. Moreover, we could categorize all the FB models into two classes from a new perspective. In the first class, the decomposition of $\mathbb{H}_{\text{c}}$ and $\mathbb{H}_{\text{tr}}$ produces different Irreps; therefore, we could directly identify $\text{Ker}(S)$. This typically happens in lattices with special geometry and isotropic hoppings, including the most prominent examples like Kagome, pyrochlore (line graph) \cite{mielke1991ferromagnetic,mielke1991ferromagnetism,mielke1992exact,bergman2008band}, Lieb (split graph) \cite{lieb1989two,ma2020spin} and Dice lattice \cite{bergman2008band}. In such circumstances, lattice is the only degree of freedom and its geometry plays an essential role in trapping the CLSs, therefore these models could be called lattice-dominant. In the second class, $\mathbb{H}_{\text{c}}$ and $\mathbb{H}_{\text{tr}}$ share the same Irreps hence symmetries alone do not guarantee a non-empty $\text{Ker}(S)$. Nevertheless, it is still possible to obtain a kernel by eliminating a particular coupling channel. Practically, we can explicitly study the hoppings. Notice that if two hopping vectors are related by an operation in $\mathcal{G}$, the corresponding matrices $t^{\langle ij\rangle}$ essentially transform to each other up to a unitary transformation. This could simplify the mathematical structure of $S$ (see Eq.~\eqref{S_honeycomb} as an example), thus assisting us in determining $\text{Ker}(S)$. It appears that the majority of high-orbital FBs belong to the latter class, indicating that the orbital degree of freedom is also indispensable. Here, although the lattice itself does not directly produce destructive interference, additional orbital degree of freedom allows the interference to equivalently occur in the internal space, thereby trapping a CLS. Such FBs could be referred to as lattice-assisted, in sharp comparison with the former class. In SM \cite{supp}, we apply our scheme to the Kagome lattice without and with orbital degree of freedom as an additional example, also demonstrating the difference between these two classes of FBs.
	
	Similar symmetry treatments are also valid in the second step. After finding a non-empty kernel,  $\text{Ker}(S)$ and $\text{Ker}^{\perp}(S)$ are also decomposed according to Irreps. Therefore, the analysis is similar, which will be clearly exemplified in the following examples.

	\section{Applications and examples}\label{sec_example}
	We now exemplify our construction scheme in various systems with different symmetries $\mathcal{G}$ and atomic orbitals. To simplify, we consider the symmetric CLS corresponding to a non-degenerate FB. It carries a one-dimensional Irrep as 
	\begin{equation}
		\label{sym}
		\ket{w(\boldsymbol{R})}\xrightarrow{g\in\mathcal{G}}\ket{w(\boldsymbol{R})}\chi_{g}
	\end{equation}
	where $\chi_{g}$ coincides with the character of the Irrep.
	
	\subsection{2D $d$-orbital honeycomb model}
	\begin{figure} [htbp]
		\centering
		\includegraphics[width= 1\columnwidth]{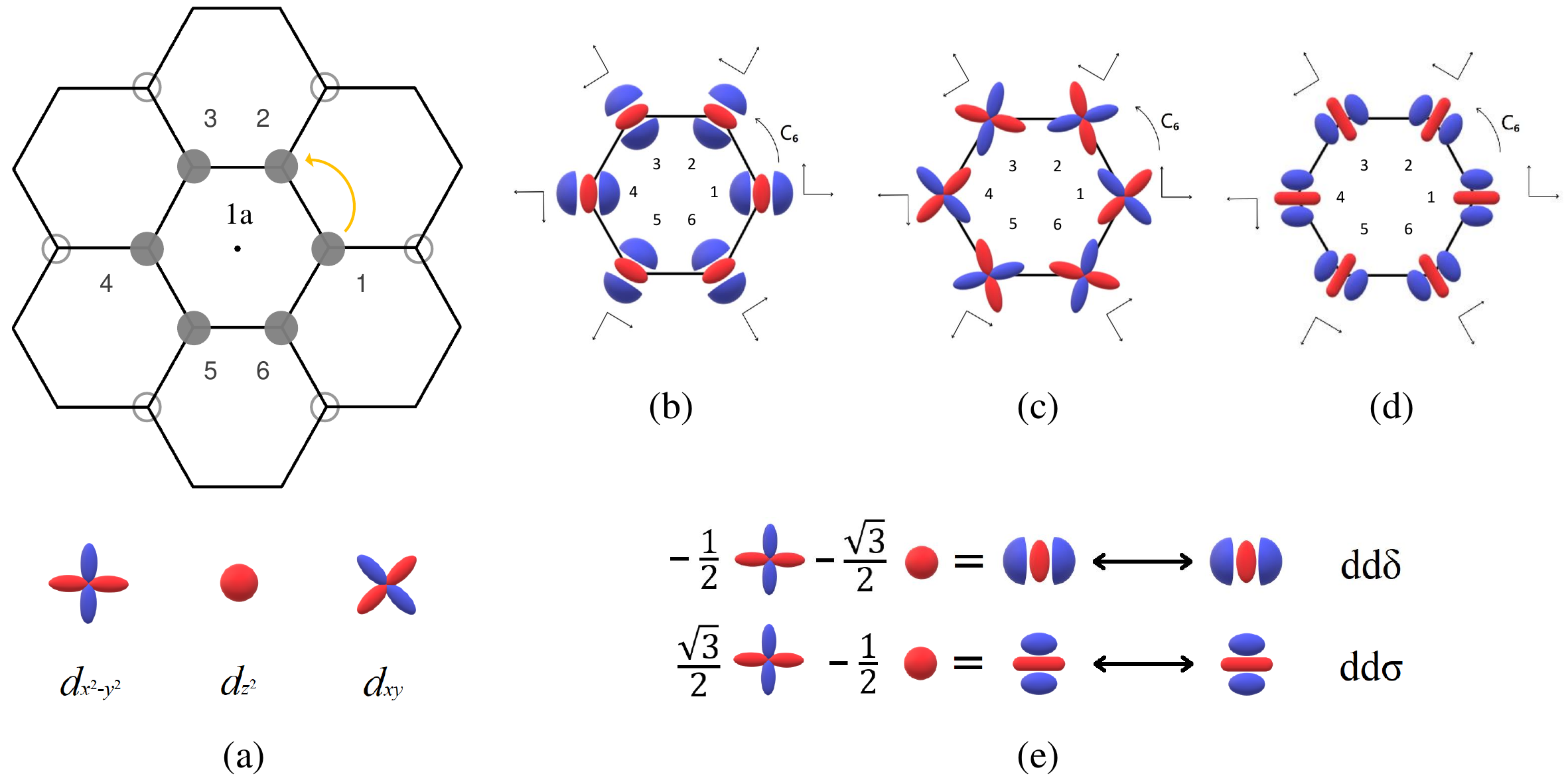}
		\caption{(a) The shape of the symmetric CLS, which centers around the 1a WP (which realizes the point group symmetry $\mathcal{G}$), and occupies the six shaded sites (2b WP). The yellow arrow indicates its generation. The CLS has six nearest neighbors marked by the circles.
			Schematic plot of three types of basis with $i$ labels the CLS sites (b) $\ket{\delta,i}$. (c) $\ket{\pi,i}$. (d) $\ket{\sigma,i}$. 
			(e) Illustration of $\ket{\delta}$ and $\ket{\sigma}$.}
		\label{d-orbital}
	\end{figure}
	
	We study the 2D honeycomb lattice with nearest-neighbor hoppings as the first example of our framework and show how the criterion Eq.~\eqref{cond} works. The lattice lacks a frustrated structure (such as the Kagome lattice) that naturally causes destructive interference and therefore is not generally recognized as a paradigm for studying FBs. Remarkably, introducing orbital degree of freedom enriches its nature, as illustrated below. The honeycomb lattice has point group $\mathcal{G}=C_{6v}$. We set the origin on the 1a WP, and the atoms occupy 2b WPs. The model has a chiral symmetry corresponding to a $\tau_{z}$ in the sublattice space, resulting in a symmetric energy spectrum. Applying our construction scheme, we start from a 2b site adjacent to the origin and perform all the elements of $\mathcal{G}$ to generate the shape of the CLS, resulting in a hexagon (Fig.~\ref{d-orbital}(a)). The CLS will only occupy these six sites. To demonstrate the generality of our scheme, we place three spinless degenerate $d_{x^{2}-y^{2}},d_{z^{2}},d_{xy}$ orbitals on each site. It is compatible with symmetry and is also common in transition-metal materials. The SOC is suppressed, i.e., $\lambda^{ij}_{\alpha\beta}=0$ in Eq.~\eqref{H}. This setting renders $\mathbb{H}_{\text{c}}$ and $\mathbb{H}_{\text{tr}}$ both 18-dimensional. Utilizing the point group $\mathcal{G}=C_{6v}$ and its character table (Table ~\ref{character}), we calculate the character of $\mathbb{H}_{\text{c}}$ and $\mathbb{H}_{\text{tr}}$ to decompose them as 
	\begin{equation}
		\label{decomposition0}
		2A_{1}\oplus 2B_{1}\oplus A_{2}\oplus B_{2}\oplus 2E_{1}\oplus 2E_{2}
	\end{equation}
	with $A,B (E)$ a one (two)-dimensional Irrep of the point group $C_{6v}$. Noticing that all the Irreps appear in the decomposition of both the two subspaces, symmetry does not guarantee a kernel and further studies on $S$ are required.
	
	\begin{table}[htbp]
		\begin{tabular}{@{}ccccccc@{}}
			\toprule
			$C_{6v}$ & E & $\text{C}_2$ & $2\text{C}_3$ & $2\text{C}_6$ & $3\sigma_d$ & $3\sigma_v$ \\ \midrule
			$A_1$ & 1 & 1  & 1   & 1    & 1  & 1           \\
			$A_2$ & 1 & 1  & 1   & 1    & $ -1 $   & $ -1 $          \\
			$B_1$ & 1 & $ -1 $   & 1    & $ -1 $   & $ -1 $   & 1           \\
			$B_2$ & 1 & $ -1 $   & 1    & $ -1 $   & 1    & $ -1 $          \\
			$E_1$ & 2 & $ -2 $   & $ -1 $ & 1   & 0    & 0           \\
			$E_2$ & 2 & 2  & $ -1 $    & $ -1 $  & 0  & 0    \\ 
			$\chi_{\text{c,tr}}$ & 18 & 0  & 0  & 0  & 0  & 2    \\ 
			\bottomrule      
		\end{tabular}\quad
		\begin{tabular}{@{}ccccccc@{}}
			\toprule
			$C_{6}$ & E & $\text{C}_6$ & $\text{C}_3$ & $\text{C}_2$ & $\text{C}^{2}_3$ & $\text{C}^{5}_6$ \\ \midrule
			$A$ & 1 & 1  & 1   & 1    & 1  & 1           \\
			$B$ & 1 & $ -1 $  & 1   & $ -1 $   & 1   & $ -1 $          \\
			$E'_{1}$ & 1 & $ \omega $  & $ \omega^{2} $ & $ -1 $  & $ \omega^{4} $ & $ \omega^{5} $  \\
			$E'_{2}$ & 1 & $ \omega^{5} $  & $ \omega^{4} $ & $ -1 $  & $ \omega^{2} $ & $ \omega $   \\
			$E''_{1}$ & 1 & $ \omega^{2} $  & $ \omega^{4} $ & $ 1 $  & $ \omega^{2} $ & $ \omega^{4}$  \\
			$E''_{2}$ & 1 & $ \omega^{4} $  & $ \omega^{2} $ & $ 1 $  & $ \omega^{4} $ & $ \omega^{2}$  \\
			$\chi_{\text{c,tr}}$ & 18 & 0  & 0  & 0  & 0  & 0    \\ 
			\bottomrule      
		\end{tabular}
		\caption{Character table of $C_{6v}$ and $C_{6}$, $\omega=\exp(i\pi/3)$. $\chi_{\text{c,tr}}$ is the character when treating the Hilbert space $\mathbb{H}_{\text{c,tr}}$ as a representation space.}
		\label{character}
	\end{table}
	
	\subsubsection{Slater-Koster formalism}
	To capture the hopping terms between high orbitals in the model, we now employ two-center SK integrals \cite{slater1954simplified} as a physical approximation. The potentials on each site are approximated as spherical, allowing us to classify the hoppings by angular momentum projected onto the hopping directions, referred to as $\sigma/\pi/\delta$-bondings for $d$-orbitals respectively. In this model, all the nearest-neighbor hoppings are related by the six-fold rotation, we refer to such hoppings related by symmetries as the same type hereafter. The hopping along $x$ direction takes
	\begin{equation}
		\label{t}
		t^{x}=
		\dfrac{1}{4}
		\begin{bmatrix}
			3dd\sigma+dd\delta & \sqrt{3}(-dd\sigma+dd\delta) & 0 \\
			\sqrt{3}(-dd\sigma+dd\delta) & dd\sigma+3dd\delta & 0 \\
			0 & 0 & 4dd\pi
		\end{bmatrix}
	\end{equation}
	where the others are written as $D^{n}(\text{C}_{6z})t^{x}D^{-n}(\text{C}_{6z})$ with $D(\text{C}_{6z})$ the representation of $\text{C}_{6z}$ in the $d$-orbital basis. The lattice structure provides a one-to-one correspondence between each CLS site and its neighbor (Fig.~\ref{d-orbital}(a)). Hence, the mapping $S$ explicitly takes a simple block-diagonal form as
	\begin{equation}\label{S_honeycomb}
		S=\mathop{\oplus}_{n=0}^{5}D^{n}(\text{C}_{6z})t^{x}D^{-n}(\text{C}_{6z}).
	\end{equation} 
	We then calculate the determinant of $S$, which reduces to the determinant of $t^{x}$ due to its block-diagonalized feature. The interference condition provides a prerequisite for obtaining a FB as $\text{Det}\left(t^{x}\right)\sim dd\delta\cdot dd\pi\cdot dd\sigma=0$. To provide an intuitive understanding of this condition, we take a specific linear combination of the $d$-orbitals as $\ket{\delta}=(-1/2)\ket{d_{x^{2}-y^{2}}}-(\sqrt{3}/2)\ket{d_{z^{2}}},\ket{\pi}=\ket{d_{xy}},\ket{\sigma}=(\sqrt{3}/2)\ket{d_{x^{2}-y^{2}}}-(1/2)\ket{d_{z^{2}}}$ based on the local rotated coordinates. Their hoppings along $x$ direction are given by $dd\delta,dd\pi,dd\sigma$, respectively. To satisfy the interference condition, the strength of a certain bonding should be negligible. Similar conditions have been termed as the block hopping scheme in a very recent study \cite{chen2023decoding}. We visualize these orbitals in Fig.~\ref{d-orbital}(b)$\sim$(e), and it is convenient to take them as new orthonormal basis on each site. Each set of the basis (Fig.~\ref{d-orbital}(b)$\sim$(d)) span a 6-dimensional subspace and we can decompose them as
	\begin{equation}
		\label{decomposition1}
		\begin{gathered}
			\text{Span}\{\ket{\delta,i}\}=A_{1}\oplus B_{1}\oplus E_{1}\oplus E_{2}\\
			\text{Span}\{\ket{\pi,i}\}=A_{2}\oplus B_{2}\oplus E_{1}\oplus E_{2}\\
			\text{Span}\{\ket{\sigma,i}\}=A_{1}\oplus B_{1}\oplus E_{1}\oplus E_{2}.
		\end{gathered}
	\end{equation}
	When $dd\delta/(dd\pi)/(dd\sigma)=0$, corresponding coupling channels in Eq.~\eqref{decomposition0} are blocked, leading to an non-empty $\text{Ker}(S)$ spanned by the wavefunctions $\ket{\delta/(\pi)/(\sigma),i}$. Notably, when only considering the lattice degree of freedom, the hopping is just a number. $\text{Ker}(S)$ is typically empty unless $t^{x}=0$, corresponding to the trivial atomic limit. In contrast, $t^{x}$ with the orbital degree of freedom (Eq.~\eqref{t}) allows for more possibilities, and the resulting FBs are lattice-assisted as we previously named. 
	
\begin{figure*} [htbp]
		\centering
		\includegraphics[width= 2\columnwidth]{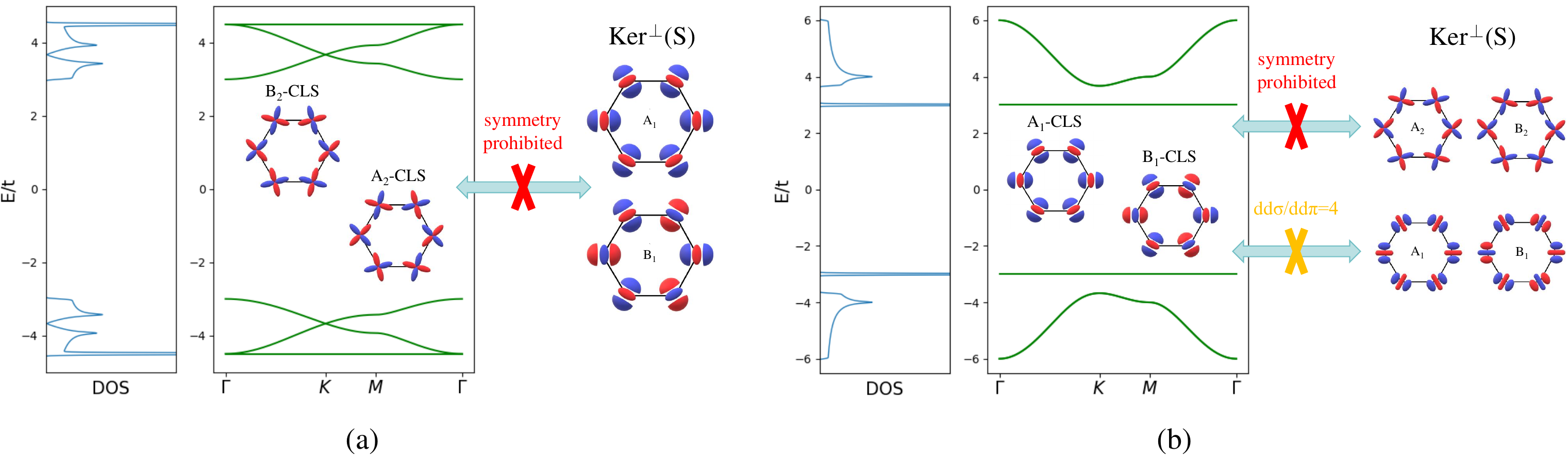}
		\caption{Band structure and DOS of the $d$-orbital honeycomb model. (a)Parameters $\boldsymbol{d}=(4,0,0)$. The $A_{2}$ and $B_{2}$ states in $\text{Ker}(S)$ are decoupled to $\text{Ker}^{\perp}(S)$, therefore manifest as the CLSs that give rise to FBs.
			(b) $\boldsymbol{d}=(4,1,0)$. The couplings within the subspace $A_{1}$ and $B_{1}$ vanish when $dd\sigma/dd\pi=4$, which leads to the FBs.}
		\label{band}
	\end{figure*}

	It is now necessary to specify the parameters to produce a non-empty $\text{Ker}(S)$ and further practical FBs. Considering the anisotropy of the high-orbitals in realistic materials, $|dd\sigma|>|dd\pi|\gg|dd\delta|$ usually holds, we set the SK integrals as $\boldsymbol{d}=(dd\sigma,dd\pi,dd\delta)=(4,0,0)$ as the first illustrative example. A common factor $t$ is disregarded without losing generality. Now $\text{Ker}(S)$ is spanned by $\ket{\delta,i}$ and $\ket{\pi,i}$. Invoking Eq.~\eqref{decomposition1}, we have
	\begin{equation}
		\begin{aligned}
			\text{Ker}(S)&=A_{1}\oplus B_{1}\oplus A_{2}\oplus B_{2}\oplus 2E_{1}\oplus 2E_{2}\\
			\text{Ker}^{\perp}(S)&=\text{Span}\{\ket{\sigma,i}\}=A_{1}\oplus B_{1}\oplus E_{1}\oplus E_{2}.
		\end{aligned}
	\end{equation}
	We focus on the one-dimensional Irreps ($A,B)$ for searching non-degenerate FBs. The next step is to distinguish eigenstates in $\text{Ker}(S)$. Manifestly, Irreps $A_{2}$ and $B_{2}$ only appear in $\text{Ker}(S)$. As long as the point group symmetry is preserved, their couplings with any states in $\text{Ker}^{\perp}(S)$ are forbidden. Therefore, they must be eigenstates and each of them is exactly the symmetric CLS in this model. The explicit form of the wavefunctions could be obtained from the projection operators defined as \cite{dresselhaus2007group}
	\begin{equation}\label{projection}
		\hat{P}^{(\Gamma^{n})}=\dfrac{d_{n}}{|\mathcal{G}|}\sum_{g\in \mathcal{G}}\chi^{(\Gamma^{n})}(g)\hat{P}_{g}.
	\end{equation}
	Here, $d_{n}$ is the dimensionality of the Irrep and $\chi$ is the character. $\hat{P}_{g}$ is the corresponding operator of $g$ acting on the state.
	After the projection, the wavefunction $\hat{P}^{(\Gamma^{n})}\ket{\varphi_{i}}$ carries the Irrep $\Gamma^{n}$. Following this step, we express the $A_{2}$-CLS  wavefunction in a unified coordinate as
	\begin{equation}\label{cls-400}
		\ket{w(\boldsymbol{R})}=\sum_{i}-\sin\theta_{i}\ket{d_{x^{2}-y^{2}},\boldsymbol{R},i}+\cos\theta_{i}\ket{d_{xy},\boldsymbol{R},i}
	\end{equation}
	where $\theta_{i}=2(i-1)\pi/3$ and each atomic orbital is located at $(\boldsymbol{R+\delta}_{i})$, with $\vec{\delta}_{i}$ defined in Eq.~\eqref{orb} and measured from the center of the CLS. The energy of the FB is calculated as $E_{\text{FB}}=-4.5$. Also, the two symmetric CLSs $(A_{2},B_{2})$ are related by the chiral symmetry operator $\tau_{z}$ and hence possess opposite energies. We plot the bands in Fig.~\ref{band}(a) to verify our analysis, as well as the CLSs. Intriguingly, the band structure and the density of states (DOS) resemble two sets of Kagome-like bands with slight deviation quantitively, while the underlying lattice is a honeycomb. This might be explained as the formation of a bond-centered Kagome lattice \cite{zhou2014sd}. To further understand the nature of these two FBs, it is worth noticing that the associated CLSs are only composed of the $d_{x^{2}-y^{2}}/d_{xy}$ doublet. They form the $A_2 (B_2)$ Irrep and their couplings with $\mathbb{H}_{\text{tr}}$ are eliminated as long as the $\pi$-bonding is negligible. Therefore, the corresponding FB will survive with $dd\sigma$ and $dd\delta$ varied, though it may not be realistic in materials. 
	
	Interestingly, $\vec{d}=(4,0,0)$ is not the only FB condition in this model according to our theory. We can relax the constraints to only $dd\delta=0$, which may mimic the parameters in real materials better. In this case, the kernel is spanned by $\ket{\delta,i}$. The decomposition yields
	\begin{equation}
		\begin{aligned}
			\text{Ker}(S)&=\text{Span}\{\ket{\delta,i}\}=A_{1}\oplus B_{1}\oplus E_{1}\oplus E_{2}\\
			\text{Ker}^{\perp}(S)&=A_{1}\oplus B_{1}\oplus A_{2}\oplus B_{2}\oplus 2E_{1}\oplus 2E_{2}.
		\end{aligned}
	\end{equation}
	We still concentrate on one-dimensional Irreps. To decouple the wavefunctions in $\text{Ker}(S)$, we investigate the hopping channels labeled by the Irrep, which formally takes
	\begin{equation}
		\label{T}
		H^{(\Gamma^{n})}=
		\begin{bmatrix}
			H_{S} & T^{\dagger} \\
			T & H^{\perp}_{S}
		\end{bmatrix}\qquad
		\Gamma^{n}=A_{1},B_{1} .
	\end{equation} 
	This is analogous to Eq.~\eqref{S}. Here, $H^{(\perp)}_{S}$ denotes the Hamiltonian of the $\Gamma^{n}$ state in the subspace $\text{Ker}^{(\perp)}(S)$ and $T$ is their coupling. Following a similar route, we denote the CLS as $\{\tilde{\psi},0\}$ in the basis $\{\text{Ker}(S),\text{Ker}^{\perp}(S)\}$. The conditions for obtaining a FB are
	\begin{equation}\label{cls}
		T\tilde{\psi}=0\qquad 
		H_{S}\tilde{\psi}=E_{\text{FB}}\tilde{\psi}
	\end{equation}
	alike to Eq.~\eqref{cond}. They might be satisfied by further fine-tuning the parameters to a proper regime. In our model, $H_{S}$ and $T$ are both numbers. Therefore, the latter in Eq.~\eqref{cls} naturally holds. To calculate $T^{(\Gamma^{n})}$, we apply $\hat{P}^{(\Gamma^{n})}$ (Eq.~\eqref{projection}) on the basis $\ket{\delta,i}$ and $\ket{\sigma,i}$ to obtain the wavefunctions (Fig.~\ref{band}(b)) and later their couplings, which is $T^{(\Gamma^{n})}=\pm 3\sqrt{3}(dd\sigma-4dd\pi)/32$, where the plus (minus) sign is for $\Gamma^{n}=A_{1} (B_{1})$. When $dd\sigma/dd\pi=4$, the coupling vanishes and two chiral FBs with the $A_1$-CLS and $B_1$-CLS are expected, which are verified in Fig.~\ref{band}(b) by setting $\boldsymbol{d}=(4,1,0)$. The wavefunction of the positive-energy $A_{1}$-CLS is given by
	\begin{widetext}
		\begin{equation}\label{cls-410}
			\ket{w(\boldsymbol{R})}=\sum_{i}-\dfrac{\cos\theta_{i}}{2}\ket{d_{x^{2}-y^{2}},\boldsymbol{R},i}-\dfrac{\sqrt{3}}{2}\ket{d_{z^{2}},\boldsymbol{R},i}-\dfrac{\sin\theta_{i}}{2}\ket{d_{xy},\boldsymbol{R},i}
		\end{equation}
	\end{widetext}
	which has non-zero components on all the three $d$-orbitals we consider. We observe that the FBs in Fig.~\ref{band}(a) touch another dispersive band at $\Gamma$, while in Fig.~\ref{band}(b) the FBs are fully-gapped. We will discuss it in detail in section \ref{sec_touching}.
	
	\begin{figure} [htbp]
		\centering
		\includegraphics[width= 1\columnwidth]{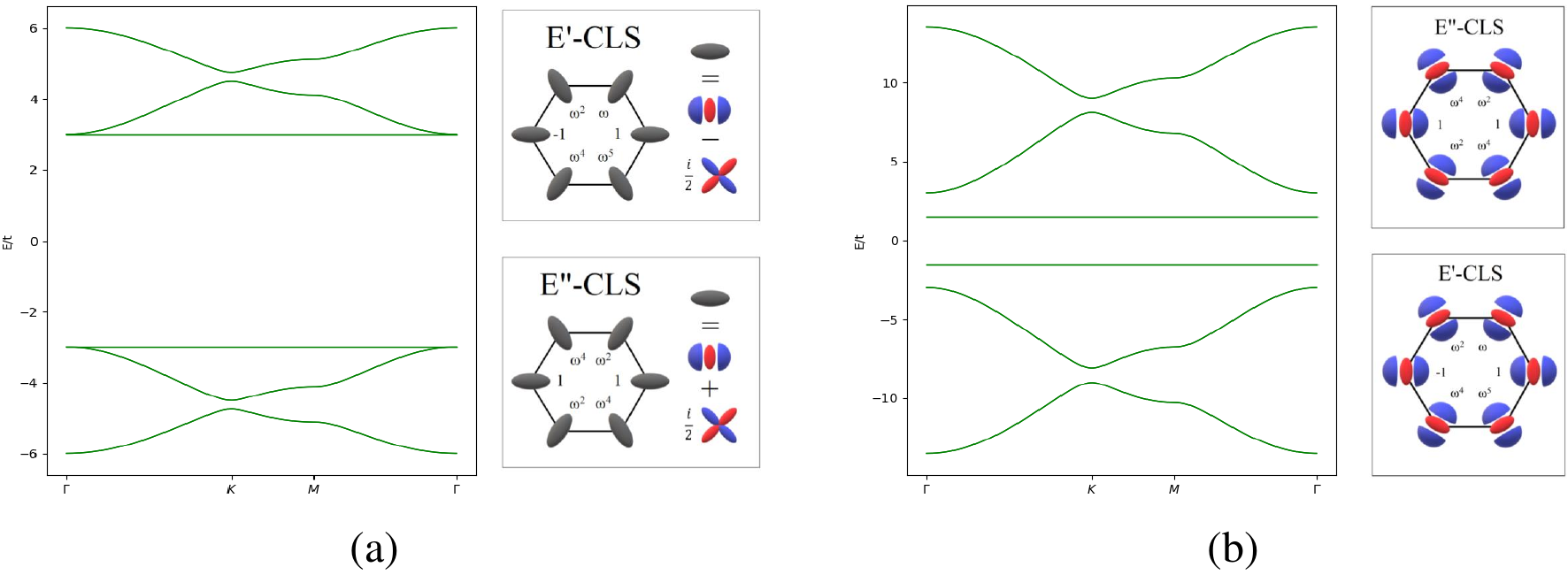}
		\caption{Band structure of the $d$-orbital honeycomb model with the SOC. The parameters are (a) $\boldsymbol{d}=(4,0,0),\lambda=\pm1.5$, (b) $\boldsymbol{d}=(4,1,0),\lambda=\pm7.5$. Assuming positive $\lambda$, the right panel shows the spin-up complex-valued CLSs.}
		\label{complex}
	\end{figure}
	
	\subsubsection{Application to the SOC}
	In realistic materials, the SOC is prevalent, particularly in heavy compounds. Extensive research have revealed that it can alter the band structure, leading to topological nontrivial bands \cite{hasan2010colloquium,bansil2016colloquium}. Nevertheless, achieving strict FBs in the presence of the SOC remains relatively underinvestigated. Remarkably, our construction scheme naturally applies when the SOC is included in real space. We double the Hilbert space of the previous model and introduce the leading on-site SOC term as
	\begin{equation}
		\bra{d_{x^{2}-y^{2}},s_{z}}\lambda\boldsymbol{L}\cdot\boldsymbol{S}\ket{d_{xy},s'_{z}}=2i\lambda s_{z} \cdot\delta_{s_{z},s'_{z}}
	\end{equation}
	which couples $d_{x^{2}-y^{2}}$ and $d_{xy}$ with the same spin on the same site. The SOC term in Eq.~\eqref{H} is reduced to $i\sum_{\alpha\beta}\sum_{i}\lambda_{\alpha\beta}c^{\dagger}_{i,\alpha}c_{i,\beta}$. The spin component $s_{z}$ is a good quantum number of the system and the two spins are related by time reversal and effectively exhibit opposite coupling strength $\lambda$. Hence, it is not necessary to involve the double-group representation. Instead, we only focus on the spin-up component, with the point group symmetry lowered from $\mathcal{G}=C_{6v}$ to $C_{6}$ (since spin is inversed under mirror as a pseudovector). Based on the new symmetry and corresponding character table (Table ~\ref{character}), the decomposition in Eq.~\eqref{decomposition1} is modified to 
	\begin{equation}
		\label{decomposition2}
		\text{Span}\{\ket{\delta(\pi)(\sigma),i}\}
		=A\oplus B\oplus E_{1}'\oplus E_{2}'\oplus E_{1}''\oplus E_{2}''.
	\end{equation}
	Here, $A,B$ denote ordinary one-dimensional Irreps of $C_{6}$ while $E_{i=1,2}' (E_{i=1,2}'')$ are one-dimensional Irreps but related by time reversal with conjugated eigenvalues. We could construct FBs by repeating the aforementioned procedure. Since the hopping terms are intact, a non-empty $\text{Ker}(S)$ still requires $dd\delta\cdot dd\pi\cdot dd\sigma=0$. Then, we investigate the Hamiltonian of each subspace $\Gamma^{n}$ (Eq.~\eqref{T}) and verify that no real-valued CLSs exist when $\lambda\neq0$. Following Eq.~\eqref{cls}, we discover FBs with the parameters tuned to
	$\boldsymbol{d}=(4,0,0),\lambda=\pm1.5$ or $\boldsymbol{d}=(4,1,0),\lambda=\pm7.5$, respectively (Fig.~\ref{complex}). The corresponding complex-valued CLSs of the spin-up sector are also shown, while the spin-down CLS can be obtained by imposing a complex conjugation. It is worth noticing that while the CLSs in Fig.~\ref{band}(a) are analogous to the early $p$-orbital honeycomb model in optical lattice \cite{wu2007flat,wu2008p}, the rest (Fig.~\ref{band}(b) and Fig.~\ref{complex}(a)(b)) are significantly distinct. This emphasizes the significance of exploiting the orbital degree of freedom and demonstrates the high-orbital system as a versatile platform for exploring FBs.\\

	\subsection{3D simple cubic lattice model}
	
	\begin{figure} [htbp]
		\centering
		\includegraphics[width= 1\columnwidth]{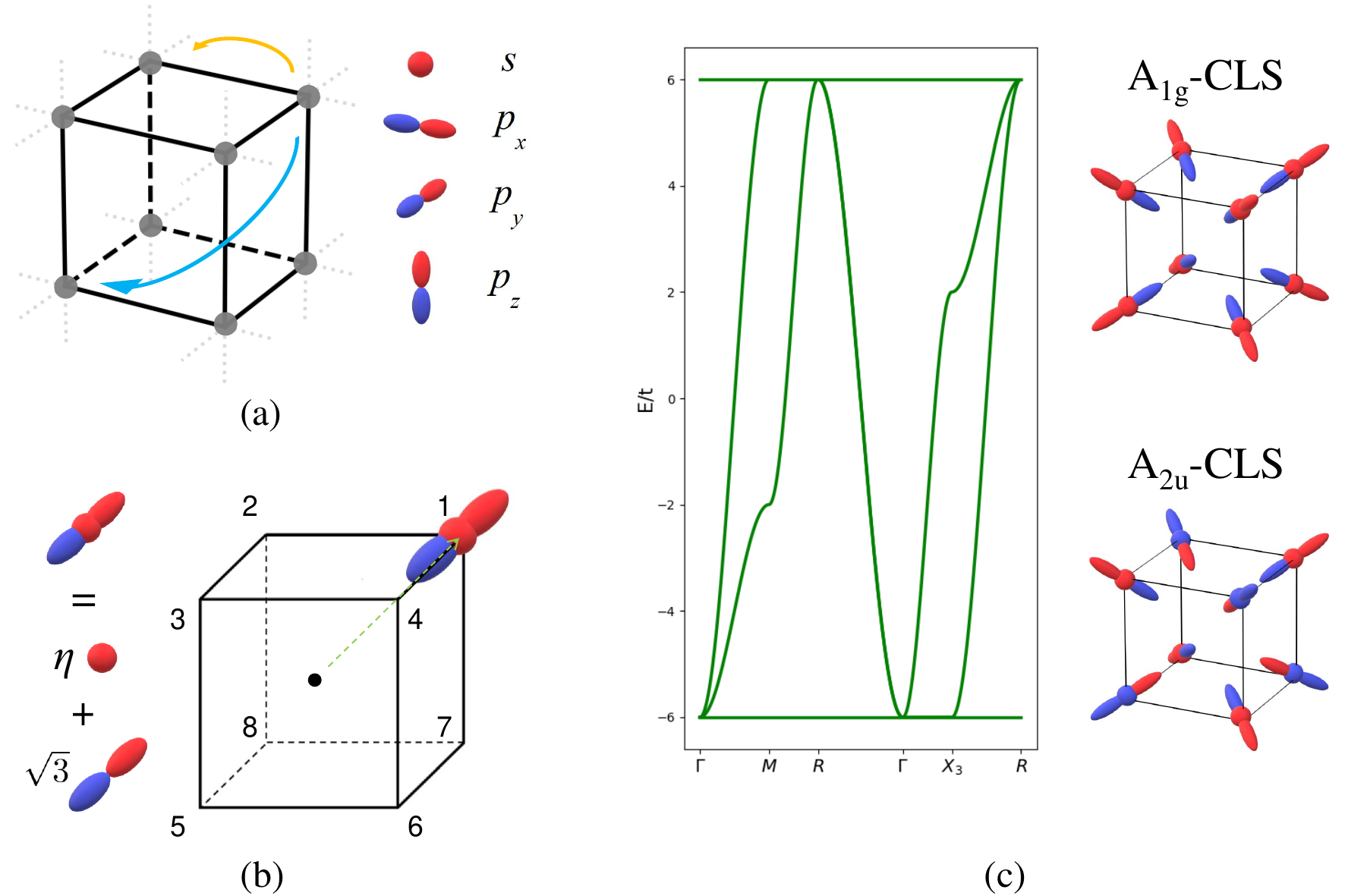}
		\caption{(a) Simple cubic lattice with $s,p_{x},p_{y},p_{z}$ orbitals. The
			yellow arrow (four-fold rotation $\text{C}_{4z}$) and blue arrow (inversion $\text{I}$) indicate the generation of the shape of the CLS.
			(b) Schematic plot of the basis $\ket{\varphi_{1},i}$ on site 1, which is a hybridization of the $s$-orbital and $p_{\boldsymbol{u}}$-orbital with $\boldsymbol{u}$ points from the center of the cubic to the corner.
			(c) Band structure and corresponding CLSs of the 3D FBs. The parameters are $ss\sigma=1,sp\sigma=\sqrt{3},pp\sigma=-3,pp\pi=0$.}
		\label{cubic}
	\end{figure}
	
	Our construction scheme could be well applied to any dimensions. In the following, we search for FBs over the whole 3D BZ in 3D lattices, which are much scarcer both theoretically and experimentally. We take a general simple cubic lattice with four degenerate $s,p_{x},p_{y},p_{z}$ orbitals residing on each site, also with nearest-neighbor hoppings. The point group of the system is $\mathcal{G}=O_{h}$. Notice that all the hopping vectors can be related by the operations in $\mathcal{G}$, the system only exhibits a unique type of hopping as well. We set the origin of the point group operations on 1a WP, i.e., the center of the cubic, and the atoms occupy 1b WPs. Starting from a neighboring site, its $\mathcal{G}$-orbit includes eight sites with a cubical shape (Fig.~\ref{cubic}(a)). Each candidate CLS site has three neighbors outside the cubic, and the corresponding mapping $S$ from $\mathbb{H}_{\text{c}}$ to $\mathbb{H}_{\text{tr}}$ diagonalizes into rectangle blocks as
	\begin{widetext}
		\begin{equation}
			S^{\text{T}}=\text{Diag}\left([t^{x},t^{y},t^{z}],[t^{\bar{x}},t^{y},t^{z}],[t^{\bar{x}},t^{\bar{y}},t^{z}],[t^{x},t^{\bar{y}},t^{z}],[t^{\bar{x}},t^{y},t^{\bar{z}}],[t^{x},t^{y},t^{\bar{z}}],[t^{x},t^{\bar{y}},t^{\bar{z}}],[t^{\bar{x}},t^{\bar{y}},t^{\bar{z}}]\right).
		\end{equation}
	\end{widetext}
	Here, we label the hopping $t$ by their direction. Specifically, $t^{x}$ is the $4\times4$ hopping matrix in $x$-direction which takes
	\begin{equation}
		t^{x}=
		\begin{bmatrix}
			ss\sigma & -sp\sigma & 0 & 0 \\
			sp\sigma & pp\sigma & 0 & 0 \\
			0 & 0 & pp\pi & 0 \\
			0 & 0 & 0 & pp\pi \\
		\end{bmatrix}.
	\end{equation}
	All other hoppings can be obtained by applying unitary transformation on $t^{x}$. Investigating each rectangle block, we find that a non-empty $\text{Ker}(S)$ demands $ss\sigma\cdot pp\sigma+sp\sigma^{2}=0$ and $ pp\pi=0$ with some algebra. Hence, we set $ss\sigma=1,sp\sigma=\eta$ and $pp\sigma=-\eta^{2}<0$ hereafter. In this regime, the kernel is 8-dimensional spanned by the wavefunctions
	\begin{equation}
		\ket{\varphi_{1},i}=C(\eta\ket{s}+\sqrt{3}\ket{p_{\boldsymbol{u}}})
	\end{equation}
	where $i$ labels the CLS sites and $C=1/\sqrt{\eta^{2}+3}$ is for normalization.
	$\boldsymbol{u}$ points from the center of the cubic to the corner (Fig.~\ref{cubic}(b)). Accordingly, we further define 
	\begin{equation}
		\begin{cases}
			\ket{\varphi_{2},i}=C(-\sqrt{3}\ket{s}+\eta\ket{p_{\boldsymbol{u}}})\\
			\ket{\varphi_{3},i}=\ket{p_{\boldsymbol{v}}}\\
			\ket{\varphi_{4},i}=\ket{p_{\boldsymbol{w}}}
		\end{cases}
	\end{equation}
	together with $\ket{\varphi_{1},i}$ to serve as a set of orthonormal basis on each site, where $\boldsymbol{u},\boldsymbol{v},\boldsymbol{w}$ are perpendicular unit vectors to guarantee orthogonality. With a non-empty $\text{Ker}(S)$, we utilize the Irreps of $O_{h}$ to perform decompositions as $\text{Ker}(S) = A_{1g} \oplus A_{2u} \oplus T_{1u} \oplus T_{2g}$ and $\text{Ker}^{\perp}(S) = A_{1g} \oplus A_{2u} \oplus E_{g} \oplus E_{u} \oplus T_{1g} \oplus 2T_{1u} \oplus 2T_{2g} \oplus T_{2u}$, with $A/E/T$ a one/two/three-dimensional Irrep of the point group $O_{h}$, respectively. Non-degenerate FBs will emerge if we could eliminate the couplings $T^{(\Gamma^{n})}$ where $\Gamma^{n}=A_{1g},A_{2u}$. To achieve that, we use $P^{(\Gamma^{n})}$ defined in Eq.~\eqref{projection} to explicitly obtain the wavefunctions. A schematic depiction of the $A_{1g}$ and $A_{2u}$ states in $\text{Ker}(S)$ is plotted in Fig.~\ref{cubic}(c). The $A_{1g}$ one has identical amplitude on the basis $\ket{\varphi_{1},i}$, resulting in an ``all-in-all-out'' orbital order; while the other $A_{2u}$ one has a staggered configuration. Their counterparts in $\text{Ker}^{\perp}(S)$ share similar characters, but on the basis $\ket{\varphi_{2},i}$. With their concrete forms, we directly calculate their couplings as $T^{(\Gamma^{n})}\propto\eta(\eta^{2}-3)$, which vanishes by tuning the parameters to $\eta=\pm\sqrt{3}$. We verify the existence of 3D FBs in Fig.~\ref{cubic}(c) by plotting the bands along the high symmetry points in the 3D BZ. Intriguingly, the FBs exhibit touchings along high-symmetry lines instead of on a single point, resulting in a nodal-line structure (see SM \cite{supp} for complementary band structures). We will provide a comprehensive analysis of them in later discussions (see section \ref{sec_touching}).
	
	\subsection{Embedding 2D CLSs in stacking structures}
	
	\begin{figure} [htbp]
		\centering
		\includegraphics[width= 1\columnwidth]{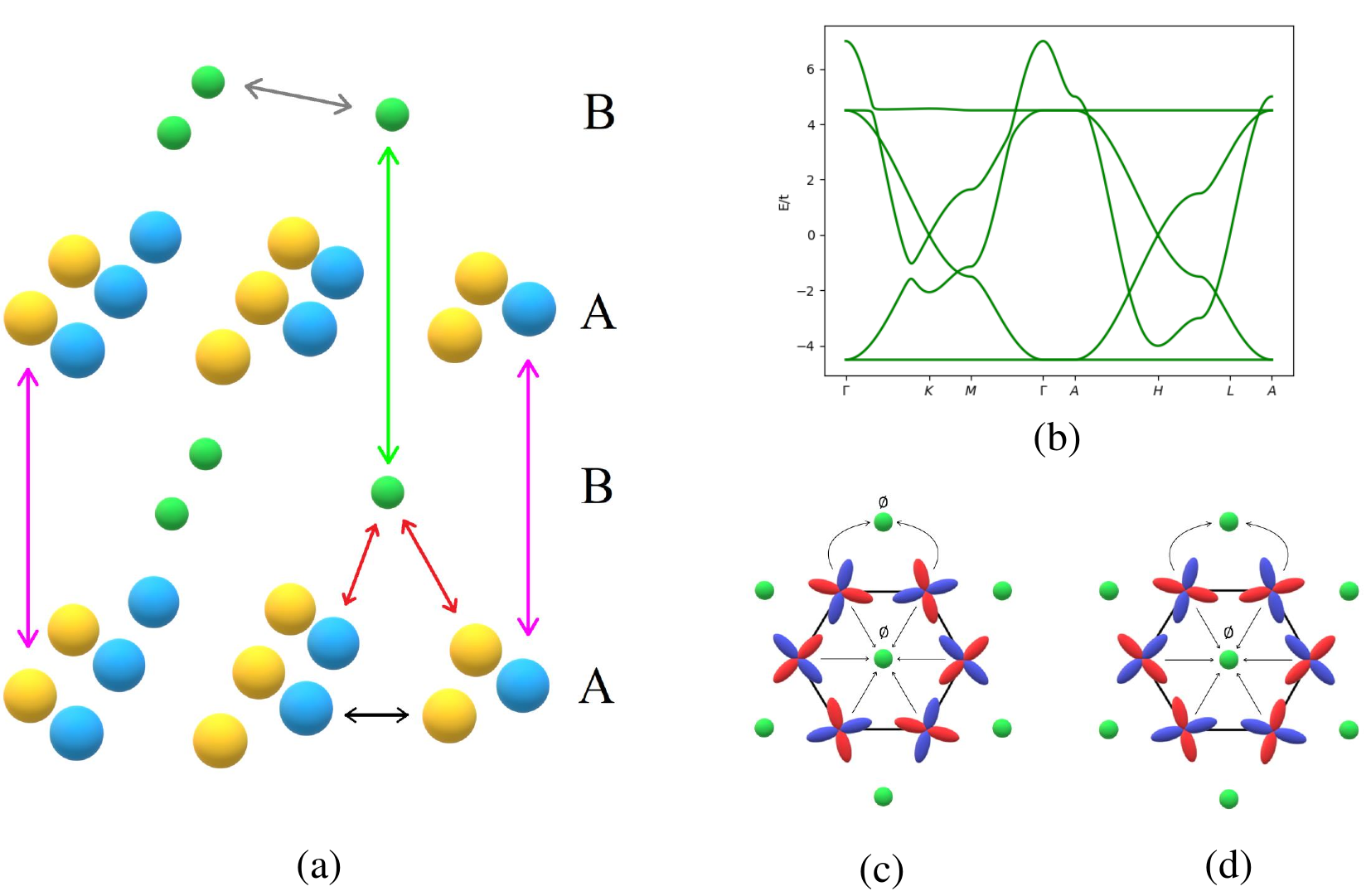}
		\caption{(a) The 3D AB stacking lattice with three atoms in one unit cell. The model exhibits five types of hoppings as indicated by black, gray (within the same A/B layer), magenta, green (between two A/B layers) and red (between A-B layers). All the layers are stacked equidistantly in calculations and the lattice constant is set as $c=2a$.
			(b) Band structure of the stacking model. The parameters are $\boldsymbol{d}=(4,0,0),ss\sigma_{xy}=1,sd\sigma=0.5,ss\sigma_{z}=0.5$. The FB exists provided that $dd\pi=dd\delta=0$ as in Fig.~\ref{band}(a).
			(c) Destructive interference (indicated by $\emptyset$) of the $A_{2}$-CLS. All the orbitals are projected onto a plane.
			(d) The original $B_{2}$-CLS. Only the hoppings to the central atoms cancel; thus, it does not strictly meet the interference condition.
		}
		\label{AB}
	\end{figure}
	
	We have utilized our method to obtain FBs in the 2D honeycomb lattice and the 3D cubic lattice, characterized by hexagonal and cubical CLSs, respectively. Meanwhile, a 3D structure can be viewed as stacking 2D lattices, especially for recently prevalent Van der Waals materials \cite{liu2016van,novoselov20162d}. We now shift the spotlight to investigate the FBs in such stacking structures. Quite remarkably, we could construct FBs in such systems by simply embedding 2D CLSs in 3D lattices. Taking the aforementioned honeycomb model as a prototype, we now exclude the $d_{z^{2}}$ orbital for simplicity, as its on-site energy may be far separated from the $d_{x^{2}-y^{2}}/d_{xy}$ doublet due to factors such as crystal field splitting. We employ this simplified four-band system to construct a 3D lattice by AB-stacking in the $z$ direction. Fig.~\ref{AB}(a) illustrates the lattice, where the A layer corresponds to the original honeycomb lattice, while the B layer consists of $s$ orbitals and aligns with the center of the honeycomb lattice, occupying a different WP and forming an in-plane triangular lattice. The point group symmetry of the system remains as $\mathcal{G}=C_{6v}$ (which may be promoted to $D_{6h}$ if the layers are stacked equidistantly, but it is not necessary for the following analysis). Five types of hoppings are included in the new model, as indicated by the colored arrows in Fig.~\ref{AB}(a). Heuristically, due to the weak inter-layer coupling nature of Van der Waals materials, it is reasonable to speculate that the system could inherit FBs from the stacking layers. Therefore, it is beneficial to revisit the 2D CLSs in Fig.~\ref{band}. If they still satisfy the general criterion (Eq.~\eqref{cond}) to manifest as a CLS in the new system, then the corresponding FB will not be disrupted. Since the $d_{z^{2}}$ orbital is excluded, we can not expect FBs with $\boldsymbol{d}=(4,1,0)$ since its CLS has $d_{z^{2}}$ component (Eq.~\eqref{cls-410}). Interestingly, one of the FBs with $\boldsymbol{d}=(4,0,0)$ does survive (Fig.~\ref{AB}(b)). To understand its nature, we first analyze the lattice structure. In the AB-stacking configuration, the $d$-orbitals in the A layer now have six new neighbors in the B layers (three from above and another three from below). We now study the influence of these new neighbors on the original 2D CLSs. For the $A_{2}$-CLS with $\chi_{\text{C}_{6z}}=1$ and $\chi_{\sigma_{d}}=-1$, it has vanishing interlayer hopping amplitude, as shown in Fig.~\ref{AB}(c). Also, the coupling between two CLS in neighboring A layers is given by $\delta$-bonding, which vanishes since $dd\delta=0$. These two properties only rely on the original symmetry group $\mathcal{G}$ and guarantee that the $A_{2}$-CLS remains an eigenstate, irrespective of the specific stacking distances in the $z$-direction. Hence, this results in a 3D FB that is characterized by the original $A_{2}$-CLS, with energy $E_{\text{FB}}=-4.5$. However, the hoppings from the original $B_{2}$-CLS to the B layer are not exactly canceled as its $\chi_{\sigma_{d}}=1$ (Fig.~\ref{AB}(d)). Therefore, the original FB with positive energy can not be completely flat when generalized to a 3D lattice (Fig.~\ref{AB}(b)). Nevertheless, since the interlayer couplings are relatively weak (we set $sd\sigma$ and $ss\sigma_{z}$ smaller than the other parameters in the calculation to reflect this feature), the bandwidth remains very narrow and still provides a strongly enhanced density of states. 
	
	We refer to the previous procedure as a (2+1)D construction, whose quintessence is to generalize a 2D FB model to a 3D one by straightforwardly utilizing the known 2D FBs together with their corresponding CLSs. Due to its symmetry and weak interlayer couplings, it could remain as an exact (or approximate) eigenstate of a new 3D model, leading to a 3D FB. While the FB models in lower dimensions have been extensively studied, (2+1)D construction could provide an intuitive real-space picture and new insights for obtaining and analyzing 3D FBs.
	
	\section{Band touching}\label{sec_touching}
	\begin{figure} [htbp]
		\centering
		\includegraphics[width= 1\columnwidth]{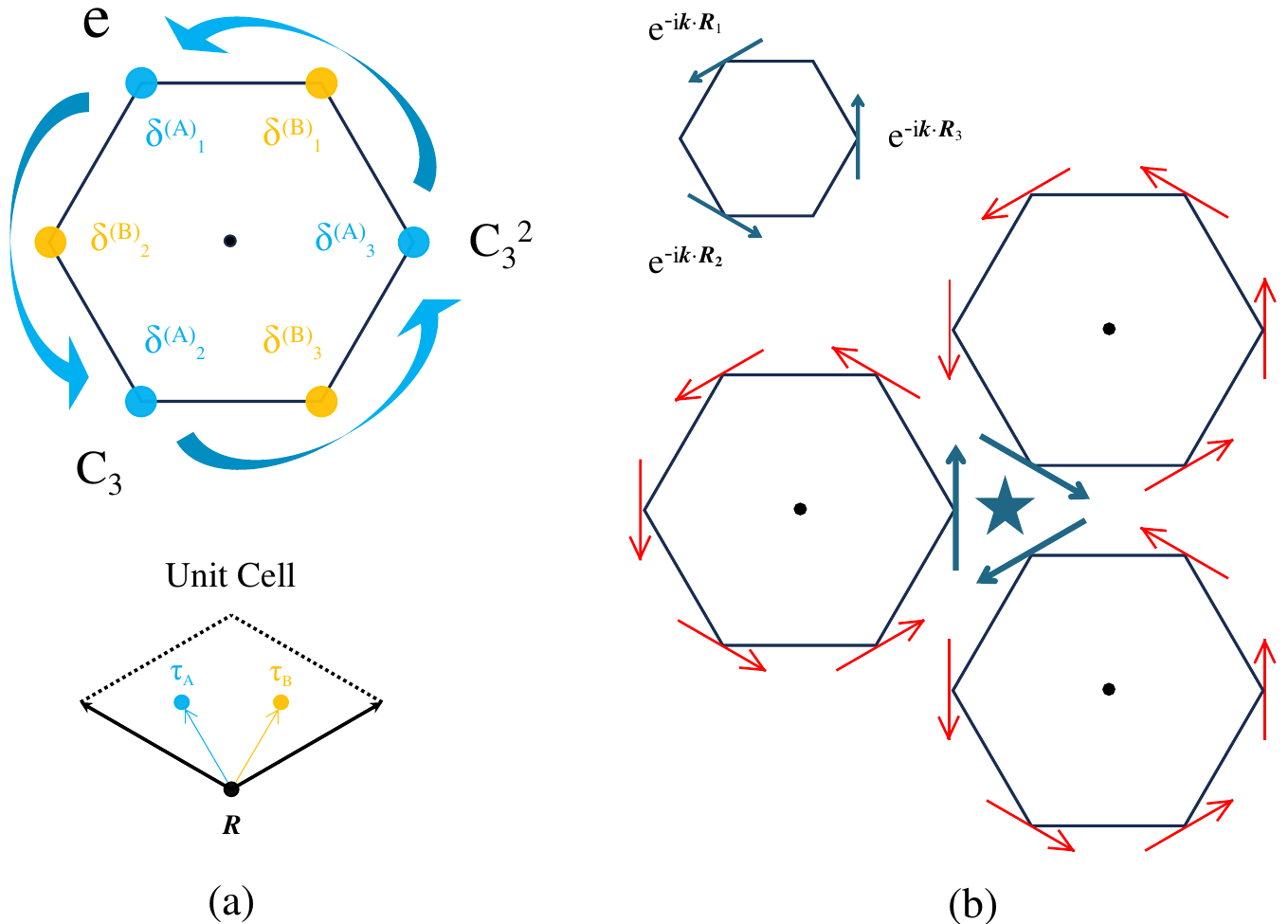}
		\caption{(a) Definition of $\vec{\delta}^{(s)}_{i}$ and $\vec{\tau}_{s}$ in the hexagonal CLS. There is a one-to-one correspondence between $\vec{\delta}^{(s)}_{i}$ and its structure group $C_{3}$ as $\vec{\delta}^{(s)}_{1}=\text{e}\vec{q}^{s},\vec{\delta}^{(s)}_{2}=\text{C}_{3}\vec{q}^{s},\vec{\delta}^{(s)}_{3}=\text{C}^{2}_{3}\vec{q}^{s}$ with $\vec{q}^{s}=\vec{\delta}^{(s)}_{1}$. The panel below shows the conventional unit cell.
			(b) Schematic plot of three adjacent CLSs. The arrows represent wavefunctions with internal orbital components and transform non-trivially under group operations. They overlap on the central site (marked by the star) but are plotted separately for clearance. 
		}
		\label{touching}
	\end{figure}
	
	\subsection{General setup}
	Our construction scheme only relies on the existence of a corresponding CLS, which is the most common feature of a FB. During the construction, we discover that a strict FB could have different band-touching behavior: no touching (gapped FB), touching point (degenerate at discrete points) and nodal-touching line (degenerate along lines). We now provide detail analysis on them for a more comprehensive understanding. Ref.~\cite{bergman2008band} first points out the general real-space origin of band touching. Notably, neighboring CLSs with finite overlap are typically non-orthogonal but still linear-independent. It becomes more subtle when investigating all the $N$ CLSs each located in the unit cell $\vec{R}$. Their linear independence could depend on the boundary condition of the lattice, termed as ``real space topology" \cite{bergman2008band}. When all CLSs are subject to some constraints under periodic boundaries, which is necessary for obtaining band structures, they may be insufficient to span the Hilbert space of the FB. To further illustrate the point, we define the Fourier transformation of the CLS in terms of its creation operator as
	\begin{equation}
		w^{\dagger}_{\boldsymbol{k}}=\sum_{\boldsymbol{R}}w^{\dagger}(\boldsymbol{R})e^{i\boldsymbol{k}\cdot\boldsymbol{R}}.
	\end{equation} 
	The states labeled by different $\vec{k}$ are orthogonal. However, they span the same space as $w^{\dagger}(\boldsymbol{R})$ do since Fourier transformation is unitary. Hence, we conclude that $\sum_{\boldsymbol{R}}w^{\dagger}(\boldsymbol{R})e^{i\boldsymbol{k}\cdot\boldsymbol{R}}\ket{0}=0$ must hold at specific $\vec{k}=\vec{k_{0}}$ when the $N$ CLSs are not linear independent. In this circumstance, the FB exhibits a singular touching at $\vec{k_{0}}$ with another dispersive band, and extra eigenstates known as the non-contractible loop states (NLSs) compensate for the incompleteness \cite{bergman2008band,rhim2019classification}. While the summation $w^{\dagger}_{\boldsymbol{k_{0}}}\ket{0}=\sum_{\boldsymbol{R}}w^{\dagger}(\boldsymbol{R})e^{i\boldsymbol{k_{0}}\cdot\boldsymbol{R}}\ket{0}$ has $N$ terms, we can simplify it by projecting onto the Hilbert space of a single site. Noticing that all the CLSs are spaced $\vec{R}$ apart from each other, it is more convenient to study their components according to sublattices. Hence, we partition the occupied sites, generated from Eq.~\eqref{orb}, as $\{\vec{\delta}_{i}\}=\cup_{s}\{\vec{\delta}^{(s)}_{i}\}$ with $s$ labels the sublattice and $i$ counts the number of the sites belonging to $s$-sublattice in a single CLS. Here, $\vec{\delta}^{(s)}_{i}$ is measured from the center of the CLS and these sites belong to different conventional unit cells, we further denote $\vec{q}^{s}=\vec{\delta}^{(s)}_{1}$ as a representative and $\vec{R}^{(s)}_{i}=\vec{\delta}^{(s)}_{i}-\vec{q}^{s}$. Thus, $\vec{\delta}^{(s)}_{i}=\vec{R}^{(s)}_{i}+\vec{\tau}_{s}$ with $\vec{\tau}_{s}$ the relative displacement of the sublattice in its home unit cell (see Fig.~\ref{touching}(a) for illustration, taking the hexagonal CLS as an example). With this definition, we divide the CLS located at $\vec{R}$ into different sublattice components as $w^{\dagger}(\boldsymbol{R})=\sum_{s}w_{s}^{\dagger}(\boldsymbol{R})$, and further expand it on the atomic orbital basis as
	\begin{equation}
		w_{s}^{\dagger}(\boldsymbol{R})
		=\sum_{i,\nu}\psi_{\nu}\left(\vec{\delta}^{(s)}_{i}\right)c_{\nu}^{\dagger}\left(\vec{R}+\vec{\delta}^{(s)}_{i}\right)
	\end{equation}
	with $\nu$ the internal index for orbital components. Since the lattice is translation-invariant, we now project $w^{\dagger}_{\boldsymbol{k_{0}}}\ket{0}=0$ onto the site in the $\vec{R}_{j}=0$ unit cell with sublattice index $s$, without losing generality. This yields
	\begin{widetext}
		\begin{equation}
			\sum_{\boldsymbol{R}}\sum_{i,\nu}
			\psi_{\nu}\left(\vec{\delta}^{(s)}_{i}\right)e^{i\boldsymbol{k_{0}}\cdot\boldsymbol{R}}
			\bra{0}c_{\mu}(\vec{R}_{j}+\vec{\tau_{s}})c_{\nu}^{\dagger}\left(\vec{R}+\vec{R}^{(s)}_{i}+\vec{\tau}_{s}\right)\ket{0}
			=\sum_{\boldsymbol{R}'}\sum_{i,\nu}
			\psi_{\nu}\left(\vec{\delta}^{(s)}_{i}\right)e^{i\boldsymbol{k_{0}}\cdot\left(\boldsymbol{R}'-\boldsymbol{R}^{(s)}_{i}\right)}
			\bra{0}c_{\mu}(\vec{\tau_{s}})c_{\nu}^{\dagger}\left(\vec{R}'+\vec{\tau}_{s}\right)\ket{0}
		\end{equation}
	\end{widetext}
	where $\vec{R}'=\vec{R}+\vec{R}^{(s)}_{i}$. This leads to
	\begin{equation}\label{touch}
		\sum_{i}
		e^{-i\boldsymbol{k_{0}}\cdot\boldsymbol{R}^{(s)}_{i}}\psi_{\mu}\left(\vec{\delta}^{(s)}_{i}\right)=0.
	\end{equation}
	We illustrate this formula in Fig.~\ref{touching}(b). When projecting $w^{\dagger}_{\boldsymbol{k_{0}}}\ket{0}$ on a particular site (marked by the star in Fig.~\ref{touching}(b)), the non-zero contributions only come from the $s$-sublattice components of several adjacent CLSs. We identify that the number of summation terms is exactly the cardinality of the set $\{\vec{\delta}^{(s)}_{i}\}$. The extra phase factors indicate that the contributions are from different CLSs that are spatially located $\vec{R}$ apart from each other. 
	
	\subsection{Structure group and touching criterion}
	The analysis above indicates that whether the touching is present is completely encoded in the CLS, and it is feasible to determine the touching point numerically via Eq.~\eqref{touch} in principle. Such touchings should be distinguished from accidental degeneracy. While the original ``real space topology" is insightful and seminal, it puts no emphasis on the symmetry that is generally acknowledged crucial for degeneracy. Building on this foundation, we find that symmetries and corresponding representations play a key role in deciphering the touching and also provide a concise criterion (see Eq.~\eqref{criterion}). This further enriches the original ``real space topology". Heuristically, we discover that we could establish a bijection between $\vec{\delta}^{(s)}_{i}$ and the group element of a point group, in this case $\mathscr{G}^{s}=C_{3}$ (Fig.~\ref{touching}(a)), i.e., $\vec{\delta}^{(s)}_{i}=O_{g_{i}}\vec{q}^{(s)},g_{i}\in\mathscr{G}^{s}$. We term $\mathscr{G}^{s}$ as the structure group of the ($s$-sublattice component of the) symmetric CLS. This observation can be made rigorous, and here we briefly sketch the main procedure. To find a one-to-one correspondence between $\{\vec{\delta}^{(s)}_{i}\}$ and the group elements of $\mathscr{G}^{s}$, the cardinality of $\{\vec{\delta}^{(s)}_{i}\}$ and the order of $\mathscr{G}^{s}$ should match. Mathematically, we can infer the cardinality using the orbit-stabilizer theorem \cite{supp}. To achieve that, we consider the following subgroup of $\mathcal{G}$
	\begin{equation}
		\mathcal{G}^{s}=\{g\in\mathcal{G}|O_{g}\vec{q}^{s}=\vec{q}^{s}+\vec{R}_{g}\}
	\end{equation}
	which enables us to treat $\{\vec{\delta}^{(s)}_{i}\}$ as a $\mathcal{G}^{s}$-orbit, i.e., $\{\vec{\delta}^{(s)}_{i}\}=\{O_{g}\vec{q}^{s}|g\in\mathcal{G}^{s}\}$. The stabilizer subgroup is defined as 
	\begin{equation}
		\text{Stab}=\{g\in\mathcal{G}^{s}|O_{g}\vec{q}^{s}=\vec{q}^{s}\}.
	\end{equation}
	Hence, $\text{Card}(\{\vec{\delta}^{(s)}_{i}\})=|\mathcal{G}^{s}|/|\text{Stab}|$ according to the orbit-stabilizer theorem.
	We first traverse $\mathcal{G}^{s}$ through all point groups (2D or 3D according to the lattice dimension). For each circumstance, spatial coordinates are classified into inequivalent WPs (of the point group $\mathcal{G}^{s}$) with different stabilizer subgroups, which are accessible on the Bilbao Crystallographic Server \cite{aroyo2011crystallography,aroyo2006bilbao1,aroyo2006bilbao2}, where the coordinates of $\{\vec{\delta}^{(s)}_{i}\}$ are also displayed after $\mathcal{G}^{s}$ and $\text{Stab}$ are specified. Therefore, it is completely geometric to find proper operations that satisfy $\vec{\delta}^{(s)}_{i}=O_{g_{i}}\vec{q}^{(s)}$, where the complexity is that the $g_{i}$ should be carefully specified to form a point group. Notably, we can greatly simplify this enumeration process by taking advantage of group structures. We perform a coset decomposition as $\mathcal{G}^{s}=\cup_{n}(g_{n}\cdot\text{Stab})$, where each coset uniquely transform $\vec{q}$ to another site \cite{supp}. This indicates that once we properly select the coset representatives $g_{n}$ such that they form a group, the structure group is found. Detailed analysis based on this idea and tabulated structure groups are provided in SM \cite{supp}. 
	
	The purpose of introducing the structure group is to convert the summation in Eq.~\eqref{touch} into the summation of group elements, which enables the application of symmetry representations and their orthogonality. Regarding the phase factor $P_{\boldsymbol{k_{0}}}(g_{i})=\exp\left(-i\boldsymbol{k_{0}}\cdot\boldsymbol{R}^{(s)}_{i}\right)=\exp\left[-i\boldsymbol{k_{0}}\cdot\left(O_{g_{i}}\boldsymbol{q}^{(s)}-\boldsymbol{q}^{(s)}\right)\right]$, they could form a representation when $\boldsymbol{k_{0}}$ is a high symmetry point. Specifically, when $O_{g_{i}}^{-1}\boldsymbol{k_{0}}=\boldsymbol{k_{0}}+\boldsymbol{G}_{i}$ for all group elements of $\mathscr{G}^{s}$, where $\boldsymbol{G}_{i}$ is a reciprocal lattice vector, we have
	\begin{equation}\label{relation1}
		P(g_{j})P(g_{i})=P(g_{j}g_{i}).
	\end{equation}
	If the symmetry of $\boldsymbol{k_{0}}$ lowers to a subgroup $\mathscr{H}^{s}\subset\mathscr{G}^{s}$. The phase factors only form a representation for $\mathscr{H}^{s}$ but not $\mathscr{G}^{s}$. Nevertheless, we can perform a coset decomposition as $\mathscr{G}^{s}=\cup_{n}\left(\mathscr{H}^{s}g_{n}\right)$ and the following relation still holds
	\begin{equation}\label{relation2}
		P(h_{j})P(g_{n})=P(h_{j}g_{n}).
	\end{equation}
	where $h_{j}\in\mathscr{H}^{s}$. The proof of these two relations is given in SM \cite{supp}, and they are crucial for deriving the touching criterion later.
	
	For a symmetric CLS corresponding to a non-degenerate FB, it is an eigenstate of any operation $g\in\mathscr{G}^{s}$. We apply $g$ on $w_{s}^{\dagger}(\boldsymbol{R}=0)$ as
	\begin{widetext}
		\begin{equation}\label{gwR}
			\hat{g}\left[w_{s}^{\dagger}(\vec{R})\right]
			=\hat{g}\sum_{i,\nu}\psi_{\nu}\left(\vec{\delta}^{(s)}_{i}\right)c_{\nu}^{\dagger}\left(\vec{\delta}^{(s)}_{i}\right)
			=\sum_{i,\nu}\psi_{\nu}\left(\vec{\delta}^{(s)}_{i}\right)\sum_{\mu}c_{\mu}^{\dagger}\left(O_{g}\vec{\delta}^{(s)}_{i}\right)D_{\mu\nu}(g)
			=\chi_{g}\sum_{i,\nu}\psi_{\nu}\left(\vec{\delta}^{(s)}_{i}\right)c_{\nu}^{\dagger}\left(\vec{\delta}^{(s)}_{i}\right)
		\end{equation}
	\end{widetext}
	Here, $g$ brings the wavefunctions at $\vec{\delta}^{(s)}_{i}$ to $O_{g}\vec{\delta}^{(s)}_{i}$, and the internal orbital components transform according to the symmetry representations. $D(g)$ is the representation matrix, which depends on what atomic orbital components the CLS has. This relation tells us that the wavefunctions on different CLS sites are related as 
	\begin{equation}\label{psi}
		\psi\left(O_{g}\vec{q}^{(s)}\right)=D'(g)\psi\left(\vec{q}^{(s)}\right)
	\end{equation}
	where we define a new representation $D'(g)=\chi^{-1}_{g}D(g)$, and $\psi$ is treated as a column vector with the orbital index suppressed hereafter. Substituting Eq.~\eqref{psi} into Eq.~\eqref{touch}, we rewrite it as
	\begin{equation}\label{criterion}
		\sum_{g\in\mathscr{G}^{s}}
		P_{\boldsymbol{k_{0}}}(g)D'(g)\psi\left(\vec{q}^{(s)}\right)=0.
	\end{equation}
	On a high symmetry point $\vec{k_{0}}$, we can treat $P(g)D'(g)$ as a direct product representation of $\mathscr{G}^{s}$. $P(g)$ only relies on $\vec{k_{0}}$ in reciprocal space, while $D'(g)$ is determined by the real space symmetric CLS, in particular, its orbital components ($D(g)$) and its own symmetry as a whole entity ($\chi_{g}$). In practice, we specify a high symmetry $\vec{k_{0}}$ and then reduce $P(g)D'(g)$ into Irreps. When the reduction does not produce the trivial Irrep, $\sum_{g\in\mathscr{G}^{s}}P(g)D'(g)=0$ and Eq.~\eqref{criterion} hold, and otherwise it is invalid. This is guaranteed by the orthogonality of Irreps \cite{supp}.
	
	\begin{table*}[htbp]
		\caption{Reduction of $D'(g)$ and $P(g)D'(g)$ according to the Irreps of $C_{4h}$.}
		\begin{tabular}{cccccc}
			\hline
			CLS &\qquad $D'(g)$ \qquad&\qquad $P_{\Gamma}(g)D'(g)$  \qquad& \qquad$P_{X_{3}}(g)D'(g)$  \qquad& \qquad$P_{M}(g)D'(g)$ \qquad& \qquad$P_{R}(g)D'(g)$ \\ \hline
			$A_{1g}$ \qquad& \qquad$A_{g}\oplus A_{u}\oplus E_{u}$\qquad& \qquad$A_{g}\oplus A_{u}\oplus E_{u}$  \qquad& \qquad$A_{u}\oplus A_{g}\oplus E_{g}$  \qquad& \qquad$B_{g}\oplus B_{u}\oplus E_{u}$  \qquad& \qquad$B_{u}\oplus B_{g}\oplus E_{g}$ \\ \hline
			$A_{2u}$ \qquad& \qquad$B_{g}\oplus B_{u}\oplus E_{g}$\qquad & \qquad$B_{g}\oplus B_{u}\oplus E_{g}$  \qquad& \qquad$B_{u}\oplus B_{g}\oplus E_{u}$  \qquad& \qquad$A_{g}\oplus A_{u}\oplus E_{g}$  \qquad& \qquad$A_{u}\oplus A_{g}\oplus E_{u}$ \\ \hline
		\end{tabular}
		\label{DP}
	\end{table*}
	
	We are now well prepared to investigate the band touchings in the FB models previously constructed. The structure group of the $d$-orbital honeycomb model is $\mathscr{G}^{s}=C_{3}$, as intuitively shown in Fig.~\ref{touching}(a). We label all its Irrep as $L_{z}=0,1,2$ according to their rotational eigenvalues. The only high symmetry points under $C_{3}$ are $\Gamma$ and $K(K')$, furnishing the following Irrep
	\begin{equation}
		P_{\Gamma}(g)=0\qquad
		P_{K}(g)=1.
	\end{equation}
	The model hosts FBs with $\vec{d}=(4,0,0)$ and $\vec{d}=(4,1,0)$. All associated CLSs share $\chi_{\text{C}_{3}}=1$ (Fig.~\ref{band}). In the former case, the symmetric CLS is given by Eq.~\eqref{cls-400} with $d_{x^{2}-y^{2}}/d_{xy}$ components. Therefore, $D'(g)=1\oplus2$ and further $P_{\Gamma}(g)D'(g)=1\oplus2,P_{K}(g)D'(g)=0\oplus2$. We conclude that the only touching point is $\Gamma$. In the latter case, the symmetric CLS is given by Eq.~\eqref{cls-410} with additional $d_{z^{2}}$ components, leading to $D'(g)=0\oplus1\oplus2$. Thus, $P_{\Gamma,K}(g)D'(g)$ contains all the Irreps of $C_{3}$ and Eq.~\eqref{criterion} does not hold and the FBs exhibit no touchings. These results only depend on the symmetric CLS obtained from the previous construction procedure, which can be confirmed before obtaining the band structures.
	
	It is more interesting to investigate the touchings in the 3D cubic model. The structure group of the cubical CLS is Abelian $\mathscr{G}^{s}=C_{4h}\sim C_{4}\times Z_{2}^{I}$, whose Irreps can be classified by eigenvalues under rotation and inversion. Its validity is verified by examining Fig.~\ref{cubic}(b), where the sites labeled by 1-8 correspond to $C_{4h}=\{\text{e},\text{C}_{4},\text{C}^{2}_{4},\text{C}^{3}_{4},\text{I},\text{IC}_{4},\text{IC}^{2}_{4},\text{IC}^{3}_{4}\}$ in order. As established, we consider the high symmetry points of $C_{4h}$, which are $\Gamma(0,0,0),M=\left(\pi/a,\pi/a,0\right),R=\left(\pi/a,\pi/a,\pi/a\right),X_{3}=\left(0,0,\pi/a\right)$, with 
	\begin{equation}
		\begin{aligned}
			&P_{\Gamma}(g)=A_{g}\qquad
			&P_{X_{3}}(g)=A_{u}\\
			&P_{M}(g)=B_{g}\qquad
			&P_{R}(g)=B_{u}.
		\end{aligned}
	\end{equation}
	The $A_{1g}/A_{2u}$-CLSs share the same orbital components ($s$ and $p_{x,y,z}$), but has their own eigenvalues of $\text{C}_{4}$ and $\text{I}$. We calculate $D'(g)$ for both of them and reduce them into Irreps after taking a direct product with $P(g)$ (Table ~\ref{DP}). According to our simple criterion, the FB characterized by the $A_{1g}$-CLS exhibits touching on $M$ and $R$, while for the $A_{2u}$ one on $\Gamma$ and $X_{3}$. It is in accordance with Fig.~\ref{cubic}(c).
	
	Despite its success, this does not account for all the touchings in the model, specifically the intriguing nodal-line touchings. On $\Gamma$-$X_{3}$ and $M$-$R$ lines, the inversion symmetry is absent. Therefore $P(g)$ can not form a representation of $\mathscr{G}^{s}=C_{4h}$ but only of $\mathscr{H}^{s}=C_{4}$. To resolve it, we move back to Eq.~\eqref{criterion} and rewrite the summation with the aid of a coset decomposition $\mathscr{G}^{s}=\cup_{n}\left(\mathscr{H}^{s}g_{n}\right)$ and Eq.~\eqref{relation2}, as 
	\begin{widetext}
		\begin{equation}\label{criterion2}
			\sum_{g\in\tilde{G}^{s}} P(g)D'(g)
			=\sum_{h\in\mathscr{H}^{s}}\sum_{g_{n}} P(hg_{n})D'(hg_{n})
			=\sum_{g_{n}}\left(\sum_{h\in\mathscr{H}^{s}}
			P(h)D'(h)\right)P(g_{n})D'(g_{n}).
		\end{equation}
	\end{widetext}
	Notice that $P(h)D'(h)$ still forms a representation of the subgroup $\mathscr{H}^{s}$, which is exactly the restricted representation $P(g)D'(g)\downarrow\mathscr{H}^{s}$. If this (reducible) representation still produces no trivial Irrep after reduction, then the summation remains vanishing. Armed with this new conclusion, we find that for the $A_{1g}$-CLS, $P_{M}(g)D'(g)\downarrow C_{4}=P_{R}(g)D'(g)\downarrow C_{4}=2B\oplus E$, supporting touchings on $M$-$R$ line. Similar analysis holds for the $A_{2u}$-CLS on $\Gamma$-$X_{3}$. On an ordinary momentum without any symmetry, $\mathscr{H}^{s}$ only has trivial Irrep and $\sum_{h}
	P(h)D'(h)\neq0$, thus touching is not protected on such $\vec{k}$. 
	
	\subsection{Summary}
	We summarize the main steps to diagnose band touchings after obtaining the symmetric CLS as follows:
	\begin{enumerate}
		\item Find the structure group $\mathscr{G}^{s}$ of the CLS, whose group element has a one-to-one correspondence with the spatial coordinate of the site. They are all tabulated in SM \cite{supp}.
		\item Consider the high symmetry points $\{\vec{k_{0}}\}$ of $\mathscr{G}^{s}$, where $P_{\boldsymbol{k_{0}}}(g_{i})=\exp\left(-i\boldsymbol{k_{0}}\cdot\boldsymbol{R}^{(s)}_{i}\right)$ furnish an Irrep.
		\item Determine the representation $D'(g)=\chi^{-1}_{g}D(g)$ according to the CLS, with $\chi$ and $D$ defined in Eq.~\eqref{sym} and Eq.~\eqref{gwR}.
		\item Reduce $P_{\boldsymbol{k_{0}}}(g)D'(g)$ into Irreps of $\mathscr{G}^{s}$. A touching presents on $\boldsymbol{k_{0}}$ if and only if the reduction does not produce the trivial Irrep.
		\item Consider a continuous path connecting different $\vec{k_{0}}$, where the symmetry is lowered to a non-trivial subgroup $\mathscr{H}^{s}$. When the restricted representation $P_{\boldsymbol{k_{0}}}(g)D'(g)\downarrow\mathscr{H}^{s}$ also produce no trivial Irreps, nodal-line touchings occur on this line.
	\end{enumerate}
	Identifying these touchings is significant for acquiring the topology of bands in 2D systems. After lifting touchings at high-symmetry points with proper perturbations, non-trivial topology could be introduced into the system, recognized as a paradigm for obtaining topological FBs \cite{rhim2019classification,herzog2024topological}. For those gapped FBs without touchings, the Chern number vanishes \cite{chen2014impossibility}. Nevertheless, they could also be novel platforms as isolated FBs have been proposed to host fragile topology \cite{chiu2020fragile,cualuguaru2022general} and quantum many-body scars \cite{kuno2020flat,kuno2021multiple} in recent studies. In 3D systems, the topological classification is different \cite{chiu2016classification} and the precise relation between touchings and band topology remains unclear. Nevertheless, discovering these nodal-line touchings itself could be of great interest.
	
	\section{Discussion and Conclusion}\label{sec_conclusion}
	In summary, we have presented a real-space construction scheme based on symmetric compact localized states (CLSs) to design flat bands (FBs) in general lattice systems. It integrates lattice and orbital degrees of freedom, providing a unified perspective and novel insights for FB systems. The CLSs naturally capture the destructive interference of the hoppings, the key feature of FBs. The quintessence of our approach is to convert the physical mechanism of destructive interference to the mathematical concept of a kernel of linear mapping. Since the tight-binding (TB) Hamiltonian and the symmetric CLS respect the lattice symmetry, we can leverage group theory to analyze the mapping and derive conditions for FBs. We mainly focus on point group symmetries in this work, which are unitary. Generalizing the framework to incorporate anti-unitary symmetries, such as magnetic groups, could be a future direction. To demonstrate the effectiveness of our method, we have successfully constructed several FB models incorporating high orbitals, also covering different dimensions and discussions on spin-orbit couplings (SOCs). Taking an alternative view, SOCs could be regarded as gauge fields that equipped hoppings with an extra phase. Therefore, our scheme has potential applications for realizing FBs in artificial gauge fields, which is commonly referred to as Aharonov-Bohm caging \cite{vidal1998aharonov,li2020non,li2022aharonov}. We also discover the presence of not only band-touching points but also lines in three-dimensional FBs. Given the importance of band touchings for topology, we employ group theory to derive a general criterion for identifying both touching points and lines. The real-space eigenstates originating from such touching lines could be an interesting topic. Their counterpart, i.e., the non-contractible loop states (NLSs) in 2D, has been experimentally observed in photonic lattices \cite{xia2018unconventional,ma2020direct}, while generalizations in 3D are yet to be explored. Our findings provide a solid foundation for designing and understanding FB systems beyond specific lattice structures, and towards their realizations in real materials.
	
	\begin{acknowledgments}
		We acknowledge useful discussions with Zheng-Xin Liu and Ying-Hai Wu. We acknowledge the support by the Innovation Program for Quantum Science and Technology (Grant No. 2021ZD0302700), the Shanghai Science and Technology Innovation Action Plan (Grant No. 24LZ1400800) and the National Natural Science Foundation of China (NSFC) (Grant No.12074133).\\
	\end{acknowledgments}	
	
	\textit{Note added:} While finalizing our manuscript, we become aware of Ref.~\cite{duan2024three}, which take steps towards the realization of FBs in 3D multi-orbital systems. Unlike their work that mainly focuses on first-principle calculations, we propose a theoretical framework for FB construction with an emphasis on obtaining a symmetric CLS systematically.

	\clearpage
	\begin{appendix}
		\begin{widetext}
			\begin{center}
				\begin{Large}
					\textbf{Supplemental Materials}
				\end{Large}
			\end{center}
			\section{Review on the tight-binding model and symmetries}
			In this section, we give a brief review of the tight-binding (TB) model and how symmetries constrain the Hamiltonian, which is implicitly adapted throughout the paper. Given a periodic lattice with $N_{\text{orb}}$ orbitals in each unit cell, we denote the Wannier basis (atomic orbital) as $\ket{\boldsymbol{R},\alpha}=c_{\alpha}^{\dagger}(\boldsymbol{R}+\boldsymbol{\tau}_{\alpha})\ket{0}$, which localizes at $(\boldsymbol{R}+\boldsymbol{\tau}_{\alpha})$ and $\alpha=1,\cdots,N_{\text{orb}}$ labels the orbitals within a unit cell. The real space TB Hamiltonian takes
			\begin{equation}
				\hat{H}=\sum_{\boldsymbol{R,R'}}\sum_{\alpha\beta}c_{\alpha}^{\dagger}(\boldsymbol{R}+\boldsymbol{\tau}_{\alpha})t^{\alpha\beta}(\boldsymbol{R-R'})c_{\beta}(\boldsymbol{R'}+\boldsymbol{\tau}_{\beta})
			\end{equation}
			where $t^{\alpha\beta}(\boldsymbol{R-R'})$ is the hopping between the $\alpha$-type orbital at $(\boldsymbol{R}+\boldsymbol{\tau}_{\alpha})$ and the $\beta$-type orbital at $(\boldsymbol{R'}+\boldsymbol{\tau}_{\beta})$. When the atomic orbitals are specified, the hoppings can be parameterized by the SK integrals, which classify the couplings into $\sigma/\pi/\delta\cdots$-bondings according to the angular momentum projected onto the hopping directions. Eq.~\eqref{H} gives an equivalent expression of the hopping by relabeling the orbitals. To convert to the $\boldsymbol{k}$-space, we employ the convention
			\begin{equation}
				c_{\boldsymbol{k},\alpha}^{\dagger}=\dfrac{1}{\sqrt{N}}\sum_{\boldsymbol{R}}e^{i\boldsymbol{k}\cdot(\boldsymbol{R}+\boldsymbol{\tau}_{\alpha})}c_{\alpha}^{\dagger}(\boldsymbol{R}+\boldsymbol{\tau}_{\alpha})\qquad
				c_{\alpha}^{\dagger}(\boldsymbol{R}+\boldsymbol{\tau}_{\alpha})=\dfrac{1}{\sqrt{N}}\sum_{\boldsymbol{k}}e^{-i\boldsymbol{k}\cdot(\boldsymbol{R}+\boldsymbol{\tau}_{\alpha})}c_{\boldsymbol{k},\alpha}^{\dagger}
			\end{equation}
			where $\ket{\boldsymbol{k},\alpha}=c_{\boldsymbol{k},\alpha}^{\dagger}\ket{0}$ is the Bloch basis. The Hamiltonian is diagonalized into $\boldsymbol{k}$-blocks as
			\begin{equation}
				\hat{H}=\sum_{\boldsymbol{k}}\sum_{\alpha\beta}c_{\boldsymbol{k},\alpha}^{\dagger}H(\boldsymbol{k})_{\alpha\beta}c_{\boldsymbol{k},\beta}
			\end{equation}
			where\begin{equation}
				H(\boldsymbol{k})_{\alpha\beta}=\sum_{\boldsymbol{R}}t^{\alpha\beta}(\boldsymbol{R})e^{-i\boldsymbol{k}\cdot(\boldsymbol{R}+\boldsymbol{\tau}_{\alpha}-\boldsymbol{\tau}_{\beta})}.
			\end{equation}
			The phase factor is directly determined by the relative displacement of the two orbitals. It should be noticed that $H(\boldsymbol{k})\neq H(\boldsymbol{k+G})$ in our convention since $c_{\boldsymbol{k},\alpha}^{\dagger}\neq c_{\boldsymbol{k+G},\alpha}^{\dagger}$, but up to a gauge transformation.\\
			
			Now we consider the spatial symmetries. A spatial transformation is defined as 
			\begin{equation}
				g_{\boldsymbol{\delta}}:\boldsymbol{r}\rightarrow O_{g}\boldsymbol{r}+\boldsymbol{\delta}
			\end{equation}
			where $O_{g}\in O(d)$ in $d$-dimension. All the spatial transformations that leave the lattice invariant form its symmetry group, which is called the space group $\mathcal{S}$ of the lattice. The multiplication of the group is defined as performing the transformations successively. The orthogonal part $O_{g}$ alone also form a group, which is called the point group $\mathcal{G}$ of the space group. It is worth noticing that $\mathcal{G}$ is not necessarily a subgroup of $\mathcal{S}$, i.e., the elements in $\mathcal{G}$ may not leave the lattice invariant. On the other hand, all the translations by Bravais-lattice (BL) vectors form a normal subgroup $\mathcal{T}$ of $\mathcal{S}$, and the quotient group $\mathcal{S}/\mathcal{T}$ is isomorphic to $\mathcal{G}$. For some space groups, it is possible to find a proper origin where $\boldsymbol{\delta}=0$ for all its group elements. They are called symmorphic where $\mathcal{G}=\mathcal{S}/\mathcal{T}$ is indeed a subgroup of $\mathcal{S}$. Otherwise, they are called non-symmorphic space groups, and at least one of the elements contains fractional translations. The actual symmetric point group that keeps the lattice invariant is typically smaller than $\mathcal{S}/\mathcal{T}$. \\
			
			In a multi-orbital system, the orbitals within a unit cell should form a representation to respect the symmetry. Specifically, if $g_{\boldsymbol{\delta}}$ transforms an $\alpha$-type orbital to another $\beta$-type orbital, then
			$O_{g}(\boldsymbol{R}+\boldsymbol{\tau}_{\alpha})+\boldsymbol{\delta}$ must be the spatial coordinate of a $\beta$-type orbital. Formally, we define a representation matrix $U_{g\boldsymbol{\delta}}$ such that
			\begin{equation}
				g_{\boldsymbol{\delta}}:c_{\alpha}^{\dagger}(\boldsymbol{R}+\boldsymbol{\tau}_{\alpha})\rightarrow
				c_{\beta}^{\dagger}( O_{g}\boldsymbol{R}+\boldsymbol{R}_{\beta\alpha}^{g\boldsymbol{\delta}}+\boldsymbol{\tau}_{\beta})\left[U_{g\boldsymbol{\delta}}\right]_{\beta\alpha}
			\end{equation}
			where $\boldsymbol{R}_{\beta\alpha}^{g\boldsymbol{\delta}}= O_{g}\boldsymbol{\tau}_{\alpha}+\boldsymbol{\delta}-\boldsymbol{\tau}_{\beta}$. If $\left[U_{g\boldsymbol{\delta}}\right]_{\beta\alpha}\neq0$, $\boldsymbol{R}_{\beta\alpha}^{g\boldsymbol{\delta}}$ must be a BL vector.\\
			
			The Bloch basis transforms as
			\begin{equation}
				\begin{aligned}
					g_{\boldsymbol{\delta}}: c_{\boldsymbol{k}, \alpha}^{\dagger} & \rightarrow \frac{1}{\sqrt{N}} \sum_{\boldsymbol{R}}  e^{i \boldsymbol{k} \cdot\left(\boldsymbol{R}+\boldsymbol{\tau}_{\alpha}\right)} c_\beta^{\dagger}\left( O_{g} \boldsymbol{R}+\boldsymbol{R}_{\beta \alpha}^{g\boldsymbol{\delta}}+\boldsymbol{\tau}_\beta\right)\left[U_{g \boldsymbol{\delta}}\right]_{\beta \alpha} \\
					& =\frac{1}{\sqrt{N}} \sum_{\boldsymbol{R}} e^{i [ O_{g}\boldsymbol{k}] \cdot [ O_{g}\left(\boldsymbol{R}+\boldsymbol{\tau}_\alpha\right)]} c_\beta^{\dagger}\left( O_{g} \boldsymbol{R}+\boldsymbol{R}_{\beta \alpha}^{g\boldsymbol{\delta}}+\boldsymbol{\tau}_\beta\right)\left[U_{g \boldsymbol{\delta}}\right]_{\beta \alpha} \\
					& =e^{-i\left( O_{g} \boldsymbol{k}\right) \cdot \boldsymbol{\delta}} \frac{1}{\sqrt{N}} \sum_{\boldsymbol{R}} e^{i\left[ O_{g} \boldsymbol{k}\right] \cdot\left[ O_{g} \boldsymbol{R}+\boldsymbol{R}_{\beta \alpha}^{g \boldsymbol{\delta}}+\boldsymbol{\tau}_{\beta}\right]} c_\beta^{\dagger}\left( O_{g} \boldsymbol{R}+\boldsymbol{R}_{\beta \alpha}^{g \boldsymbol{\delta}}+\boldsymbol{\tau}_\beta\right)\left[U_{g \boldsymbol{\delta}}\right]_{\beta \alpha} \\
					& =e^{-i\left( O_{g} \boldsymbol{k}\right) \cdot \boldsymbol{\delta}} \frac{1}{\sqrt{N}} \sum_{\boldsymbol{R}'} e^{i\left[ O_{g} \boldsymbol{k}\right] \cdot\left[\boldsymbol{R}'+\boldsymbol{\tau}_\beta\right]} c_\beta^{\dagger}\left(\boldsymbol{R}'+\boldsymbol{\tau}_\beta\right)\left[U_{g \boldsymbol{\delta}}\right]_{\beta \alpha} \\
					& =e^{-i\left( O_{g} \boldsymbol{k}\right) \cdot \boldsymbol{\delta}} c^{\dagger}_{ O_{g} \boldsymbol{k},\beta}\left[U_{g \boldsymbol{\delta}}\right]_{\beta\alpha}
				\end{aligned}
			\end{equation}
			where $\boldsymbol{R'}= O_{g}\boldsymbol{R}+\boldsymbol{R}_{\beta\alpha}^{g\boldsymbol{\delta}}$. The factor $e^{-i\left( O_{g} \boldsymbol{k}\right) \cdot \boldsymbol{\delta}}$ encodes the effect of the fractional translation of non-symmorphic space group. We could define
			\begin{equation}
				\left[\hat{g}_{\boldsymbol{\delta}}(\boldsymbol{k})\right]_{\beta\alpha}
				=e^{-i\left( O_{g} \boldsymbol{k}\right)\cdot\boldsymbol{\delta}}\left[U_{g \boldsymbol{\delta}}\right]_{\beta\alpha},
			\end{equation}
			therefore $g_{\boldsymbol{\delta}}:c_{\boldsymbol{k}, \alpha}^{\dagger} \rightarrow 
			c_{ O_{g}\boldsymbol{k}, \beta}^{\dagger}\left[\hat{g}_{\boldsymbol{\delta}}(\boldsymbol{k})\right]_{\beta\alpha}$. In the main text, we only consider symmorphic space group and hence the orthogonal part $O_{g}$ of the transformation determines all the properties.\\
			
			If the Hamiltonian is invariant, i.e.,
			\begin{equation}
				g_{\boldsymbol{\delta}}: \hat{H}=\sum_{\boldsymbol{k}}\sum_{\alpha\beta}c_{\boldsymbol{k},\alpha}^{\dagger}H(\boldsymbol{k})_{\alpha\beta}c_{\boldsymbol{k},\beta}\rightarrow \hat{H},
			\end{equation}
			then
			\begin{equation}
				\hat{g}_{\boldsymbol{\delta}}(\boldsymbol{k})H(\boldsymbol{k})\hat{g}^{\dagger}_{\boldsymbol{\delta}}(\boldsymbol{k})=H(O_{g}\boldsymbol{k}).
			\end{equation}
			Now we consider the $\boldsymbol{k}$-space eigenstates with the creation operator $a^{\dagger}_{\boldsymbol{k}}=\sum_{\alpha}v_{\boldsymbol{k}}(\alpha)c_{\boldsymbol{k},\alpha}^{\dagger}$ (with the band index suppressed), which satisfies
			\begin{equation}
				\sum_{\beta}H(\boldsymbol{k})_{\alpha\beta}v_{\boldsymbol{k}}(\beta)=E_{\boldsymbol{k}}v_{\boldsymbol{k}}(\alpha).
			\end{equation}
			And we have
			\begin{equation}
				\begin{cases}
					H(\boldsymbol{k})v_{\boldsymbol{k}}=E_{\boldsymbol{k}}v_{\boldsymbol{k}}\\
					H( O_{g}\boldsymbol{k})v_{ O_{g}\boldsymbol{k}}=E_{ O_{g}\boldsymbol{k}}v_{ O_{g}\boldsymbol{k}}
				\end{cases}\qquad
				\hat{g}_{\boldsymbol{\delta}}(\boldsymbol{k})H(\boldsymbol{k})\hat{g}^{\dagger}_{\boldsymbol{\delta}}(\boldsymbol{k})=H( O_{g}\boldsymbol{k}).
			\end{equation}
			This leads to $v_{ O_{g}\boldsymbol{k}}=\hat{g}_{\boldsymbol{\delta}}(\boldsymbol{k})v_{\boldsymbol{k}}$ and $E_{ O_{g}\boldsymbol{k}}=E_{\boldsymbol{k}}$, which indicates that different $\vec{k}$ points in the Brillouin zone that can be connected via point group elements share the same eigenvalues. The $\boldsymbol{k}$-space eigenstate transforms as
			\begin{equation}
				g_{\boldsymbol{\delta}}: a^{\dagger}_{\boldsymbol{k}}=\sum_{\alpha}v_{\boldsymbol{k}}(\alpha)c_{\boldsymbol{k},\alpha}^{\dagger}\rightarrow
				\sum_{\alpha\beta\gamma}\left[\hat{g}^{\dagger}_{\boldsymbol{\delta}}(\boldsymbol{k})\right]_{\alpha\gamma}v_{ O_{g}\boldsymbol{k}}(\gamma)c_{ O_{g}\boldsymbol{k},\beta}^{\dagger}\left[\hat{g}_{\boldsymbol{\delta}}(\boldsymbol{k})\right]_{\beta\alpha}
				=\sum_{\beta}v_{ O_{g}\boldsymbol{k}}(\beta)c_{ O_{g}\boldsymbol{k},\beta}^{\dagger}
				=a^{\dagger}_{ O_{g}\boldsymbol{k}}.
			\end{equation}
			
			\section{Kagome model}
			
			\begin{figure*} [htbp]
				\centering
				\includegraphics[width= 1\columnwidth]{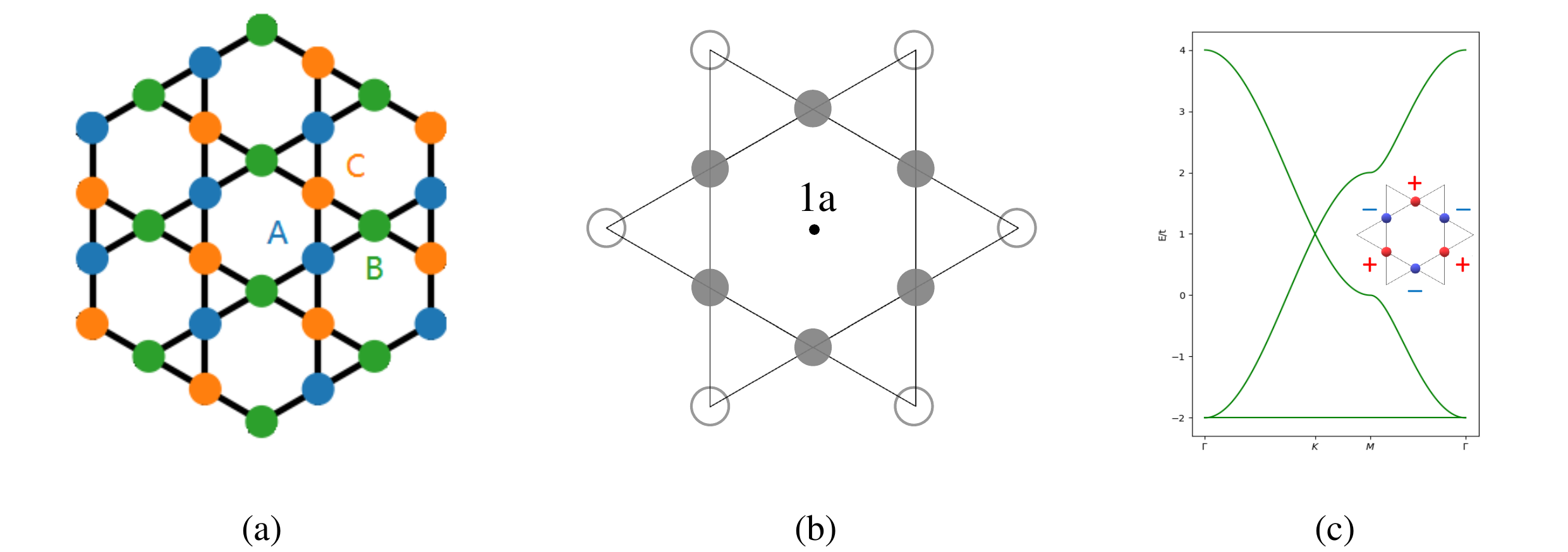}
				\caption{(a) Lattice structure of the Kagome lattice. Colors denote three sublattices A,B,C.
					(b) The shape of the symmetric CLS which occupies the six shaded sites. The yellow arrow indicates its generation. The CLS has six nearest neighbors marked by the circles.
					(c) Band structure of the three-band $s$-orbital Kagome model. The inset shows the symmetric $B_{1}$-CLS associated to the FB.}
				\label{kagome}
			\end{figure*}
			
			In this section, we apply our construction scheme to the Kagome lattice, serving as complementary examples and demonstrating the significance of symmetry in our scheme. The Kagome lattice is a prominent FB model due to its special structure. The lattice is known as the line graph of the honeycomb lattice, and the presence of FB is proven in the graph theory community. Instead of the technical details, our construction offers a physically intuitive and self-consistent understanding of its FB. The Kagome lattice shares the same symmetry group $\mathcal{G}=C_{6v}$ with the honeycomb lattice, but the atoms occupy 3c WP leading to three sublattices (Fig.~\ref{kagome}(a)). We first consider the simplest spinless $s$-orbitals with identical nearest-neighbor hoppings and follow the same FB construction procedure as described in the main text. Firstly, we specify the shape of the symmetric CLS as a hexagon by generating a $\mathcal{G}$-orbit (Eq.~\eqref{orb}), as illustrated in Fig.~\ref{kagome}(b). Then, the mapping $S$ from $\mathbb{H}_{\text{c}}$ to $\mathbb{H}_{\text{tr}}$ can be represented as a $6\times6$ matrix. Remarkably, when decomposing these two subspaces using the Irreps of $C_{6v}$, we obtain
			\begin{equation}
				\mathbb{H}_{\text{c}}=A_{1}\oplus B_{1}\oplus E_{1}\oplus E_{2}\qquad
				\mathbb{H}_{\text{tr}}=A_{1}\oplus B_{2}\oplus E_{1}\oplus E_{2}.
			\end{equation}
			Geometrically, the sites belonging to $\mathbb{H}_{\text{c}}$ or $\mathbb{H}_{\text{tr}}$ both form hexagons, but with a  $90{}^{\circ}$ rotation relative to each other due to the structure of the Kagome lattice. This leads to the difference when decomposing then by the Irreps and suggests that the $B_{1}$-wavefunction is decoupled from any other states, without any further calculations. In our terminology, we have
			\begin{equation}
				\text{Ker}(S)=B_{1}\qquad
				\text{Ker}^{\perp}(S)=A_{1}\oplus E_{1}\oplus E_{2}.
			\end{equation} 
			Therefore, we conclude that the model will host a FB (whose energy is $-2$, derived from $H\ket{w(\boldsymbol{R})}=E_{\text{FB}}\ket{w(\boldsymbol{R})}$) characterized by a $B_{1}$-CLS. We plot them in Fig.~\ref{kagome}(c).
			
			\begin{figure*} [htbp]
				\centering
				\includegraphics[width= 1\columnwidth]{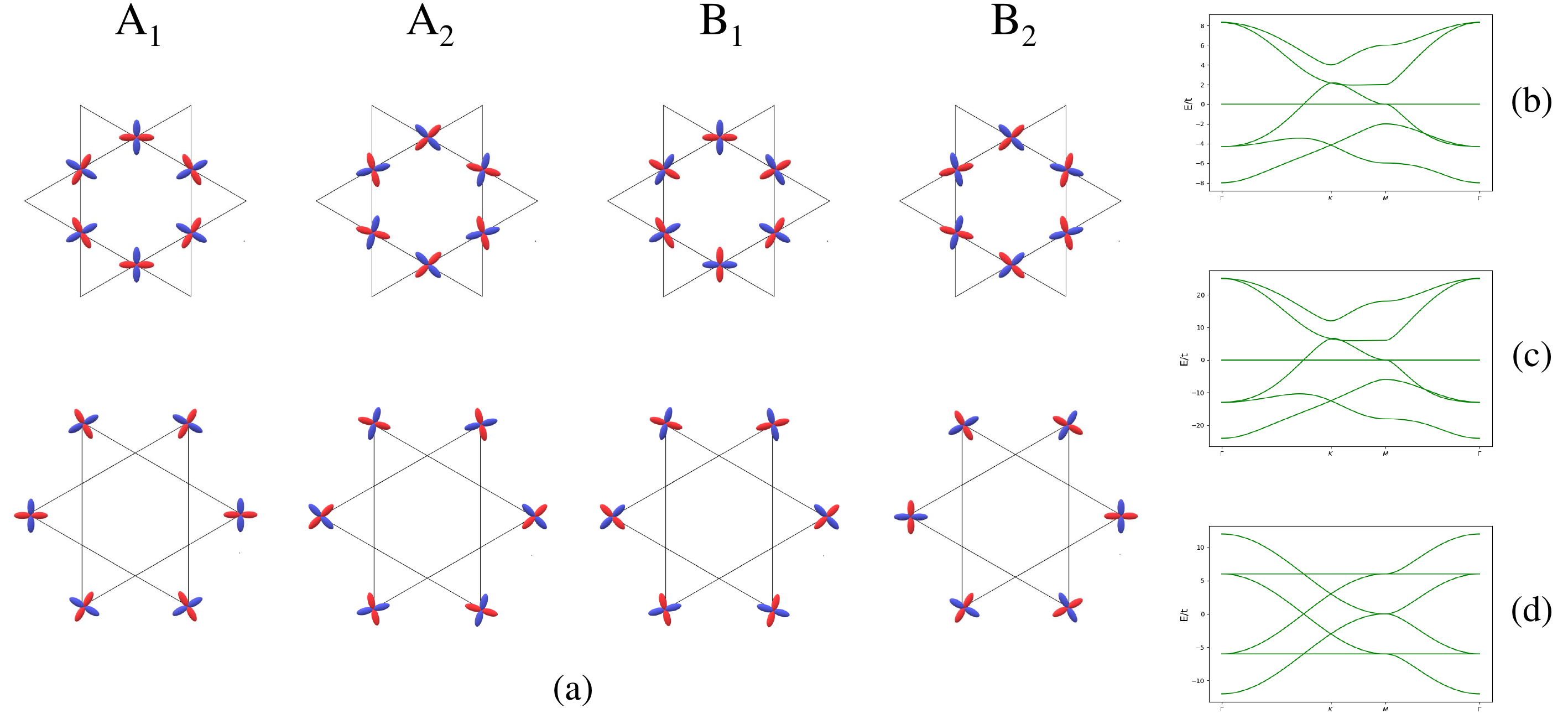}
				\caption{(a) Wavefunctions associated with the one-dimensional Irrep. Upper and lower ones belong to $\mathbb{H}_{\text{c}}$ and $\mathbb{H}_{\text{tr}}$, respectively. (b)(c)(d) Band structure with parameters 
					b) $\vec{d}=(4,1,0)$ with $A_{1}$-CLS;
					c) $\vec{d}=(4,9,0)$ with $A_{2}$-CLS;
					d) $\vec{d}=(4,-3,0)$ with $B_{1}$-CLS and $B_{2}$-CLS.}
				\label{wf}
			\end{figure*}
			
			We further consider Kagome lattice with $d$-orbitals, which are common in practical Kagome materials. Intriguingly, anisotropic hoppings from high orbitals may disrupt original destructive interference. Based on our scheme, we take the $d_{x^{2}-y^{2}}/d_{xy}$ doublet as an example and parameterize the hoppings with the SK integrals $\vec{d}=(dd\sigma,dd\pi,dd\delta)$. The candidate CLS sites are still chosen as shown in Fig.~\ref{kagome}(b), and the decomposition takes
			\begin{equation}
				\mathbb{H}_{\text{c}}=\mathbb{H}_{\text{tr}}=
				A_{1}\oplus A_{2}\oplus B_{1}\oplus B_{2}\oplus 2E_{1}\oplus 2E_{2}.
			\end{equation}
			This is distinct from the previous analysis without orbital degree of freedom. In Fig.~\ref{wf}(a), we explicitly plot all the wavefunctions associated with the one-dimensional Irreps $A,B$ to provide an intuitive understanding. The mapping $S$ is classified into four channels labeled as $A_{1},A_{2},B_{1},B_{2}$. Their coupling, which is purely a real number, can be directly calculated with the knowledge of the concrete wavefunctions. FBs are present only when certain channel is blocked, and the results are summarized in Table \ref{dkagome}. We choose proper parameters to plot the band structures in Fig.~\ref{wf}(b)(c)(d) to confirm the previous analysis. It is clear that high-orbital FBs in Fig.~\ref{wf}(b)(c) are distinct from the original band structure in Fig.~\ref{kagome}(c).
			
			\begin{table}[htbp]
				\caption{The FB condition for the Kagome lattice with $d_{x^{2}-y^{2}}/d_{xy}$ doublet. The first row denotes the symmetric CLS that characterized the FB, and the second row shows the corresponding parameters in terms of SK integrals.}
				\begin{tabular}{ccccc}
					\hline
					Symmetric CLS & $A_{1}$ & $A_{2}$ & $B_{1}$ & $B_{2}$ \\ \hline
					FB condition  
					& $\hspace{0.35em} 3dd\sigma-12dd\pi+dd\delta=0
					\hspace{0.35em}$ 
					& $\hspace{0.35em} 9dd\sigma-4dd\pi+3dd\delta=0
					\hspace{0.35em}$ 
					& $\hspace{0.35em} 3dd\sigma+4dd\pi+dd\delta=0
					\hspace{0.35em}$ 
					& $\hspace{0.35em} 3dd\sigma+4dd\pi+dd\delta=0
					\hspace{0.35em}$ \\ \hline
				\end{tabular}
				\label{dkagome}
			\end{table}

			\section{Complementary band structures of the 3D cubic lattice model}
			In this section, we provide complementary band structures of the 3D cubic lattice model to better show its nodal-line touchings of the FB. In Fig.~\ref{cubic}(c), the high symmetry points are labeled as conventional as
			\begin{equation}
				\Gamma(0,0,0)\qquad
				M(\pi/a,\pi/a,0)\qquad
				R(\pi/a,\pi/a,\pi/a)\qquad
				X_{3}(0,0,\pi/a).
			\end{equation}
			The upper FB with $A_{1g}$-CLS exhibit a touching line on $M$-$R$, while the lower one with $A_{2u}$-CLS on $\Gamma$-$X_{3}$. The point group of the system is $\mathcal{G}=O_{h}$. The symmetry constraints that $E_{O_{g}\boldsymbol{k}}=E_{\boldsymbol{k}},\forall g\in\mathcal{G}$. Therefore, we can infer that the upper FB exhibit touchings on all the edges of the BZ (Fig.~\ref{nodaltouching}(a)), and the lower FB exhibit touchings on all the $\vec{k}$-axis (Fig.~\ref{nodaltouching}(b)). We further illustrate them in Fig.~\ref{nodaltouching}(c) by choosing an alternative path in the BZ, with 
			\begin{equation}
				M'(0,\pi/a,\pi/a)\qquad
				X_{2}(0,\pi/a,0).
			\end{equation}
			
			\begin{figure*} [htbp]
				\centering
				\includegraphics[width= 0.7\columnwidth]{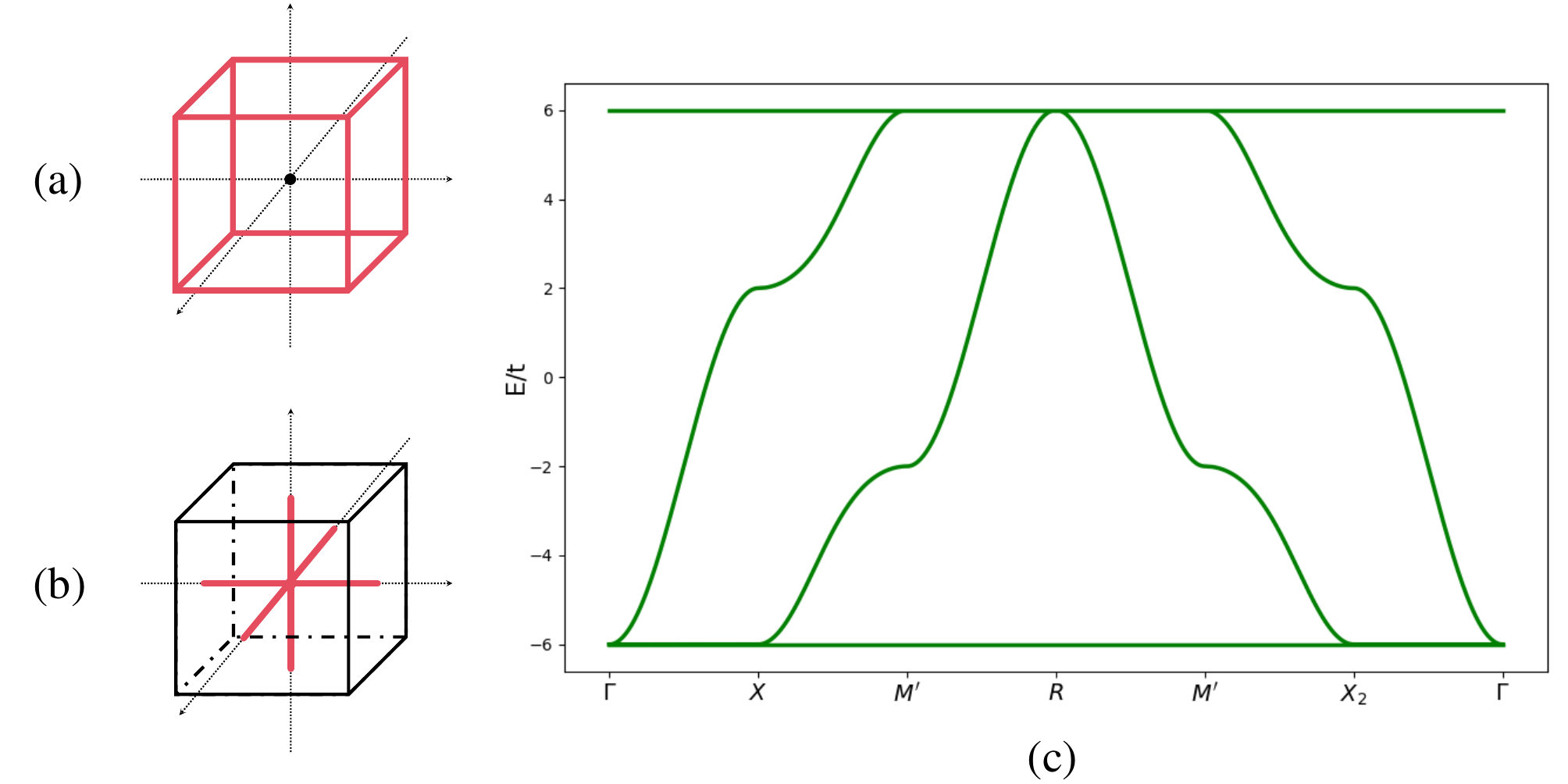}
				\caption{(a)(b) Illustration of nodal touchings of the 3D cubic lattice model. 
					a) Upper FB with $A_{1g}$-CLS;
					b) Lower FB with $A_{2u}$-CLS;
					(c) Complementary band structures of the 3D cubic lattice model.}
				\label{nodaltouching}
			\end{figure*}

			\section{Orbit and orbit-stabilizer theorem}
			In this section, we provide basic mathematical backgrounds about orbit and orbit-stabilizer theorem in group theory. Formally, we first define a (left) group action $f$ of $G$ on $X$ as a function
			\begin{equation}
				f:G\times X\rightarrow X
			\end{equation}
			where $G$ is a group and $X$ is a set. This function should satisfy 
			\begin{equation}
				f_{\text{e}}(x)=x\qquad f_{g}(f_{h}(x))=f_{gh}(x)
			\end{equation}
			where $\text{e}$ is the identity of the group. $X$ is called a $G$-set. Intuitively, a group action induces a permutation on the elements in $X$ that respect the multiplication of the group $G$. We may denote the group action as $gx$ directly for simplicity.\\
			
			Consider an element $x\in X$ with $X$ a $G$-set, the orbit of $x$ is the set of elements that can be moved by the elements of $G$, i.e.,
			\begin{equation}
				\text{Orb}_{x}=\{gx|g\in G\},
			\end{equation}
			which can be regarded as an equivalent class.\\
			
			The stabilizer subgroup of $x$ is defined as
			\begin{equation}
				\text{Stab}_{x}=\{g\in G|gx=x\}.
			\end{equation}
			
			We can perform a (left) coset decomposition of $G$ according to the stabilizer subgroup, denoted as 
			\begin{equation}
				G=\cup_{n=1}^{N}(g_{n}\cdot\text{Stab}_{x}).
			\end{equation}
			Here, $N=|G|/|G_{x}|$. $g_{n}$ are the representatives of the coset and $g_{1}=\text{e}$. Since $\text{Stab}_{x}$ leave $x$ invariant, each coset ($n\neq 1$) moves $x$ to $g_{n}x\neq x$. For two distinct cosets, their coset representatives satisfy $g_{m}^{-1}g_{n}\notin\text{Stab}_{x}$, therefore $g_{m}^{-1}g_{n}x\neq x$ and $g_{m}x\neq g_{n}x$. Thus, we conclude that each coset uniquely moves $x$ to another $x'$, and 
			the cardinality of $\text{Orb}_{x}$ is exactly $N$. This is referred to as the orbit-stabilizer theorem.

			\section{Proof of Eq.~\eqref{relation1} and Eq.~\eqref{relation2}}
			In this section, we prove Eq.~\eqref{relation1} and Eq.~\eqref{relation2} in the main text. Considering the phase factor
			\begin{equation}
				P_{\boldsymbol{k_{0}}}(g_{i})=\exp\left(-i\boldsymbol{k_{0}}\cdot\boldsymbol{R}^{(s)}_{i}\right)=\exp\left[-i\boldsymbol{k_{0}}\cdot\left(O_{g_{i}}\boldsymbol{q}^{(s)}-\boldsymbol{q}^{(s)}\right)\right]
			\end{equation}
			where $\boldsymbol{R}^{(s)}_{i}=O_{g_{i}}\boldsymbol{q}^{(s)}-\boldsymbol{q}^{(s)}$ is by definition and $\boldsymbol{k_{0}}$ satisfies $O_{g_{i}}^{-1}\boldsymbol{k_{0}}=\boldsymbol{k_{0}}+\boldsymbol{G}_{i}$ for all group elements of $\mathscr{G}^{s}$. We omit some subscripts and superscripts for simplicity and rewrite it as 
			\begin{equation}
				P(g_{i})=\exp\left(-iO_{g_{i}}^{-1}\boldsymbol{k_{0}}\cdot\boldsymbol{q}\right)
				\exp\left(i\boldsymbol{k_{0}}\cdot\boldsymbol{q}\right)
				=\exp\left(-i\boldsymbol{G}_{i}\cdot\boldsymbol{q}\right).
			\end{equation}
			It follows that
			\begin{equation}
				P(g_{j})P(g_{i})=
				\exp\left[-i(\boldsymbol{G}_{j}+\boldsymbol{G}_{i})\cdot\boldsymbol{q}\right]
			\end{equation}
			and\begin{equation}
				P(g_{j}g_{i})
				=\exp\left(-i\boldsymbol{k_{0}}\cdot O_{g_{j}}O_{g_{i}}\boldsymbol{q}\right)\exp\left(i\boldsymbol{k_{0}}\cdot\boldsymbol{q}\right)
				=\exp\left(-iO^{-1}_{g_{i}}O^{-1}_{g_{j}}\boldsymbol{k_{0}}\cdot\boldsymbol{q}\right)\exp\left(i\boldsymbol{k_{0}}\cdot\boldsymbol{q}\right)
				=\exp\left[-i(O^{-1}_{g_{i}}\boldsymbol{G}_{j}+\boldsymbol{G}_{i})\cdot\boldsymbol{q}\right].
			\end{equation}
			They are equal since
			\begin{equation}
				\dfrac{P(g_{j})P(g_{i})}{P(g_{j}g_{i})}
				=\exp\left[i\boldsymbol{G}_{j}\cdot(O_{g_{i}}\boldsymbol{q}-\boldsymbol{q})\right]
				=\exp\left(i\boldsymbol{G}_{j}\cdot\boldsymbol{R}_{i}\right)
				=1.
			\end{equation}
			
			Now consider that $O_{h_{j}}^{-1}\boldsymbol{k_{0}}=\boldsymbol{k_{0}}+\boldsymbol{G}_{j}$ only holds for a subgroup $\mathscr{H}^{s}\subset\mathscr{G}^{s}$. We perform a (right) coset decomposition as $\mathscr{G}^{s}=\cup_{n}\left(\mathscr{H}^{s}g_{n}\right)$ with $g_{n}$ referred to as the coset representatives and $g_{1}=e$. For $g_{i}\in\mathscr{H}^{s}g_{n\neq1}$, $P(g_{i})$ does not form a representation. Nevertheless, $P(h_{j})$ is a representation of $\mathscr{H}^{s}$, and $O_{h_{j}}^{-1}\boldsymbol{k_{0}}=\boldsymbol{k_{0}}+\boldsymbol{G}_{h_{j}},O_{g_{n}}^{-1}\boldsymbol{k_{0}}=\boldsymbol{k_{0}}+\boldsymbol{\Delta}_{g_{n}}$ where $\boldsymbol{\Delta}_{g_{n}}$ is not a reciprocal lattice vector. This gives
			\begin{equation}
				P(h_{j})P(g_{n})=e^{-i\boldsymbol{k_{0}}\cdot\boldsymbol{R}_{h_{j}}}e^{-i\boldsymbol{k_{0}}\cdot\boldsymbol{R}_{g_{n}}}
				=e^{-i\boldsymbol{k_{0}}\cdot\left(O_{h_{j}}\boldsymbol{q}-\boldsymbol{q}\right)} e^{-i\boldsymbol{k_{0}}\cdot\left(O_{g_{n}}\boldsymbol{q}-\boldsymbol{q}\right)}
				=e^{-i\boldsymbol{G}_{h_{j}}\cdot\boldsymbol{q}}e^{-i\boldsymbol{\Delta}_{g_{n}}\cdot\boldsymbol{q}}.
			\end{equation}
			On the other hand,
			\begin{equation}
				P(h_{j}g_{n})
				=e^{-i\boldsymbol{k_{0}}\cdot\left(O_{h_{j}}O_{g_{n}}\boldsymbol{q}-\boldsymbol{q}\right)} 
				=e^{-iO_{g_{n}}^{-1}O_{h_{j}}^{-1}\boldsymbol{k_{0}}\cdot\boldsymbol{q}}e^{i\boldsymbol{k_{0}}\cdot\boldsymbol{q}}
				=e^{-iO^{-1}_{g_{n}}\boldsymbol{G}_{h_{j}}\cdot\boldsymbol{q}}e^{-i\boldsymbol{\Delta}_{g_{n}}\cdot\boldsymbol{q}}
				=e^{-i\boldsymbol{G}_{h_{j}}\cdot O_{g_{n}}\boldsymbol{q}}e^{-i\boldsymbol{\Delta}_{g_{n}}\cdot\boldsymbol{q}}.
			\end{equation}
			Therefore 
			\begin{equation}
				\dfrac{P(h_{j})P(g_{n})}{P(h_{j}g_{n})}
				=\exp\left[i\boldsymbol{G}_{h_{j}}\cdot(O_{g_{n}}\boldsymbol{q}-\boldsymbol{q})\right]
				=\exp\left(i\boldsymbol{G}_{h_{j}}\cdot\boldsymbol{R}_{g_{n}}\right)
				=1.
			\end{equation}
			and $P(h_{j}g_{n})=P(h_{j})P(g_{n})$ holds.

			\section{Detailed analysis on Eq.~\eqref{criterion}}
			In this section, we perform a detailed analysis of Eq.~\eqref{criterion} in the main text. We begin by proving the following lemma
			\begin{equation}\label{lemma}
				\sum_{g\in G}\Gamma(g)=
				\begin{cases}
					|G|\quad \text{trivial Irrep}\\
					0\qquad \text{others}
				\end{cases}
			\end{equation}
			with $\Gamma(g)$ a Irrep of the finite group $G$. The character is defined as 
			\begin{equation}
				\chi(g)=\text{Tr}[\Gamma(g)]
			\end{equation}
			Now we consider
			\begin{equation}
				\Gamma^{-1}(h)\left(\sum_{g\in G}\Gamma(g)\right)\Gamma(h)
				=\sum_{g\in G}\Gamma(h^{-1}gh)=\sum_{g\in G}\Gamma(g)
			\end{equation}
			where $h$ is any group element of $G$ and the rearrangement theorem is used. According to Schur's lemma, we have
			\begin{equation}
				\left[\sum_{g\in G}\Gamma(g),\Gamma(h)\right]=0\qquad
				\sum_{g\in G}\Gamma(g)=\lambda I.
			\end{equation}
			According to the orthogonal theorem for characters of Irreps, we know
			\begin{equation}
				\dfrac{1}{|G|}\sum_{g\in G}
				\chi^{(i)}(g)\bar{\chi}^{(j)}(g)=\delta^{ij}
			\end{equation} 
			where $i,j$ labels the Irreps. We take $\Gamma^{(j)}$ as the trivial Irrep, then
			\begin{equation}
				\sum_{g\in G}\chi(g)=
				\begin{cases}
					|G|\quad \text{trivial Irrep}\\
					0\qquad \text{others}
				\end{cases}.
			\end{equation}
			Since 
			\begin{equation}
				\sum_{g\in G}\chi(g)
				=\sum_{g\in G}\text{Tr}[\Gamma(g)]
				=\text{Tr}\left[\sum_{g\in G}\Gamma(g)\right]
				=\lambda,
			\end{equation}
			the lemma (Eq.~\eqref{lemma}) is proven.\\
			
			In Eq.~\eqref{criterion}, $\psi\left(\vec{q}^{(s)}\right)$ is a vector with non-zero components. $P_{\boldsymbol{k_{0}}}(g)D'(g)$ form a (typically reducible) representation of $\mathscr{G}^{s}$ when $\boldsymbol{k_{0}}$ is a high symmetry point. According to Eq.~\eqref{lemma}, we knows $\sum_{g\in\mathscr{G}^{s}}P_{\boldsymbol{k_{0}}}(g)D'(g)$ is similar to the diagonal matrices $\text{Diag}\left(|G|,\cdots,0,\cdots\right)$, where $|G|$ appears $n$ times with $n$ the number of trivial Irreps after reducing $P_{\boldsymbol{k_{0}}}(g)D'(g)$, and the rest elements are 0. Hence,
			\begin{equation}
				\sum_{g\in\mathscr{G}^{s}}
				P_{\boldsymbol{k_{0}}}(g)D'(g)\psi\left(\vec{q}^{(s)}\right)=0
			\end{equation}
			only holds when the reduction of $P_{\boldsymbol{k_{0}}}(g)D'(g)$
			does not produce the trivial Irrep.\\
			
			When the symmetry of $\boldsymbol{k_{0}}$ lowers to a non-trivial subgroup $\mathscr{H}^{s}\subset\mathscr{G}^{s}$. We replace the summation in Eq.~\eqref{criterion} by Eq.~\eqref{criterion2}, which is
			\begin{equation}
				\sum_{g\in\tilde{G}^{s}} P(g)D'(g)
				=\sum_{h\in\mathscr{H}^{s}}\sum_{g_{n}} P(hg_{n})D'(hg_{n})
				=\sum_{g_{n}}\left(\sum_{h\in\mathscr{H}^{s}}
				P(h)D'(h)\right)P(g_{n})D'(g_{n}).
			\end{equation}
			If $P_{\boldsymbol{k_{0}}}(g)D'(g)\downarrow\mathscr{H}^{s}$ also produces no trivial Irreps, $\sum_{h}
			P(h)D'(h)=0$ and Eq.~\eqref{criterion} is still true. For an ordinary $\boldsymbol{k_{0}}$ without any symmetry, i.e., $\mathscr{H}^{s}=\text{e}$ is trivial, this leads to $\sum_{h}
			P(h)D'(h)\neq0$. For rigorous, we point out that Eq.~\eqref{criterion} may hold even $\sum_{h}P(h)D'(h)$ does not vanish. However, such peculiar FB models are yet to be discovered, to the best of our knowledge. Also, such ``touching" is considered irrelevant to symmetry and sensitive to perturbation. Thus, it is not of general interest.

			\section{Enumeration of structure groups}
			In this section, we explain how to find all the structure groups for all inequivalent lattice structures and tabulate them. This serves as a solid foundation for determining the band touchings of the FB.  
			In our construction procedure, the occupied sites of the symmetric CLS are defined as a $\mathcal{G}$-orbit as
			\begin{equation}
				\text{Orb}_{\boldsymbol{q}}=\{O_{g}\vec{q}|g\in\mathcal{G}\}
			\end{equation}
			where $\vec{q}$ is a site and $\mathcal{G}$ is the point group of the system. Concretely, we apply every group element of $\mathcal{G}$ on $\vec{q}$. This generates several distinct sites and we denote them as $\{\vec{\delta}_{i}\}$. Since the orbit is an equivalent class, we can reselect $\vec{q}$ as any $\vec{\delta}_{i}$, generating the same orbit. Notably, these sites could belong to different sublattices. We now choose $\vec{q}=\vec{q}^{s}$ with $\vec{q}^{s}\in\{\vec{\delta}_{i}\}$ and belong to the $s$-sublattice. We define
			\begin{equation}
				\mathcal{G}^{s}=\{g\in\mathcal{G}|O_{g}\vec{q}^{s}=\vec{q}^{s}+\vec{R}_{g}\}.
			\end{equation}
			It is straightforward to verify that $\mathcal{G}^{s}$ is a subgroup of $\mathcal{G}$, and the orbit $\{\vec{\delta}^{(s)}_{i}\}=\{O_{g}\vec{q}^{s}|g\in\mathcal{G}^{s}\}$ is a subset of $\text{Orb}_{\boldsymbol{q}}$.
			
			In our 2D honeycomb model, it can be recognized that $\mathcal{G}=C_{6v}$ and $\mathcal{G}^{s}=C_{3v}$ and in the 3D cubic model, both $\mathcal{G}=\mathcal{G}^{s}=O_{h}$ since it only has one sublattice.
			
			\begin{figure*} [htbp]
				\centering
				\includegraphics[width= 1\columnwidth]{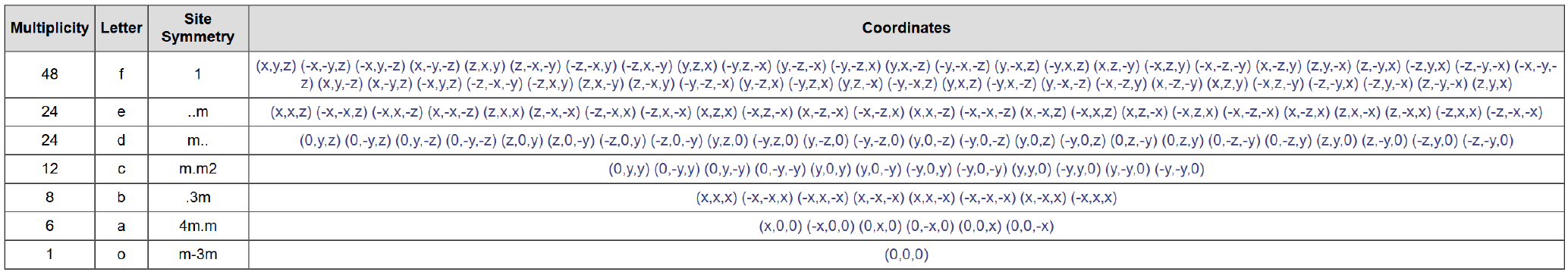}
				\caption{WPs of the 3D crystallographic point group O(432), adapted from the Bilbao Crystallographic Server (``3D WYCKPOS'' program).}
				\label{WP}
			\end{figure*}
			
			When $\mathcal{G}^{s}$ is determined, spatial coordinates are classified into inequivalent WPs. Such classification essentially depends on the stabilizer subgroup of the site (usually referred to as the site-symmetry group in crystallography). In our case, it is defined as 
			\begin{equation}
				\text{Stab}=\{g\in\mathcal{G}^{s}|O_{g}\vec{q}^{s}=\vec{q}^{s}\}.
			\end{equation}
			It is worth pointing out that here WPs are in the sense of a point group. This distinguishes it from what is commonly referred to, which is based on a space group. The case $\mathcal{G}^{s}=\text{Stab}$ is excluded since it results in an orbit with only one site, contradicting our construction scheme. We take the 3D cubic model with $\mathcal{G}^{s}=O_{h}$ as an example. All the related information is accessible on the Bilbao Crystallographic Server, which is shown in Fig.~\ref{WP}. We obtain the structure group by investigating the coset decomposition
			\begin{equation}
				\mathcal{G}^{s}=\cup_{n}(g_{n}\cdot\text{Stab}).
			\end{equation}
			From the orbit-stabilizer theorem, we know that if $g_{n}$ form a point group, then this is exactly the structure group. To achieve that, we first give a brief review of point groups. We classify 32 3D point groups into three types. Type-I only contains rotations
			\begin{equation}
				C_{n}\quad n=1,2,3,4,6\qquad
				D_{n}\quad n=2,3,4,6\qquad
				T \qquad
				O
			\end{equation}
			
			We take the direct product of type-I point groups and inversion ($Z_{2}^{I}$) to get type-II point groups
			\begin{equation}
				\begin{gathered}
					C_{1}\rightarrow C_{i}\qquad
					C_{2}\rightarrow C_{2h}\qquad
					C_{3}\rightarrow C_{3i}\qquad
					C_{4}\rightarrow C_{4h}\qquad
					C_{6}\rightarrow C_{6h}\\
					D_{2}\rightarrow D_{2h}\qquad
					D_{3}\rightarrow D_{3d}\qquad
					D_{4}\rightarrow D_{4h}\qquad
					D_{6}\rightarrow D_{6h}\\
					T\rightarrow T_{h}\qquad
					O\rightarrow O_{h}
				\end{gathered}
			\end{equation}
			
			To construct type-III point groups, we choose a type-I point group that is homomorphic to $Z_{2}$ with kernel denoted as $G_{0}$. The group $G'=G_{0}+I(G-G_{0})$ is a point group isomorphic to $G$. They are listed as
			\begin{equation}
				\begin{gathered}
					C_{2}(C_{1})\cong C_{s}\qquad
					C_{4}(C_{2})\cong S_{4}\qquad
					C_{6}(C_{3})\cong C_{3h}\\
					D_{2}(C_{2})\cong C_{2v}\qquad
					D_{3}(C_{3})\cong C_{3v}\qquad
					D_{4}(C_{4})\cong C_{4v}\qquad
					D_{6}(C_{6})\cong C_{6v}\qquad
					D_{4}(D_{2})\cong D_{2d}\qquad
					D_{6}(D_{3})\cong D_{3h}\\
					O(T)\cong T_{d}
				\end{gathered}
			\end{equation}
			where the kernel is shown in the bracket.\\
			
			Considering a group $G$ with a semidirect-product structure as $G=G_{1}\rtimes G_{2}$ or $G=G_{2}\rtimes G_{1}$, if we perform coset decomposition on $G$ with respect to $G_{1}$, then all the coset representatives can be chosen as the elements in $G_{2}$. This conclusion greatly reduces the workload of enumeration. For instance, it helps us to resolve all the structure groups of $D_{n}=C_{n}\rtimes Z_{2}$. Also, we get the structure groups of type-II point groups after those for type-I point groups are found, due to their direct-product structure. For a few exceptional cases, we can get all the coordinates of $\{\vec{\delta}^{(s)}_{i}\}$ on the Bilbao Crystallographic Server and investigate them case by case.
			
			In our 3D cubic model, the CLS sites are on the 8b WPs (we again emphasize that it is the sense of point group, as we state that it is on the 1b WPs in the main text, when in the sense of corresponding space group). We notice that the structure group may not be unique. In this case, one can choose it as $C_{4h}$ of $D_{4}$. Both of them fulfill the requirement of a structure group. However, it is always convenient to choose the simpler Abelian one. 
			
			All the structure groups are tabulated below. They are also applicable for simpler 2D point groups with proper modification.
			
			\newpage
			\begin{table}[h!]
				\caption{Tabulated structure groups for all 3D point groups.}
				\begin{tabular}{ccccccccccc}
					\hline
					\multicolumn{2}{c}{$C_{1}(1)$} &  & \multicolumn{2}{c}{$C_{i}\left(\bar{1}\right)$} &  & \multicolumn{2}{c}{$C_{2}(2)$} &  & \multicolumn{2}{c}{$C_{s}(m)$} \\\hline
					WP      & \quad Structure Group \quad      &  & WP               & \quad Structure Group \quad              &  & WP      & \quad Structure Group \quad      &  & WP      & \quad Structure Group \quad      \\\hline
					1a      & $e$                  &  & 2a               & $C_{i}$                      &  & 2b      & $C_{2}$              &  & 2b      & $C_{s}$              \\
					&                      &  & 1o               & $e$                          &  & 1a      & $e$                  &  & 1a      & $e$   
					\\\hline              
				\end{tabular}
			\end{table}
			
			\begin{table}[h!]
				\begin{tabular}{ccccccccccc}
					\hline
					\multicolumn{2}{c}{$C_{2h}(2/m)$} &  & \multicolumn{2}{c}{$D_{2}(222)$} &  & \multicolumn{2}{c}{$C_{2v}(mm2)$} &  & \multicolumn{2}{c}{$D_{2h}(mmm)$} \\\hline
					WP        & \quad Structure Group \quad       &  & WP       & \quad Structure Group \quad       &  & WP        & \quad Structure Group \quad       &  & WP        & \quad Structure Group \quad       \\\hline
					4c        & $C_{2h}$               &  & 4d       & $D_{2}$               &  & 4d        & $C_{2v}$              &  & 8g        & $D_{2h}$              \\
					2b        & $C_{2}$               &  & 2c       & $C_{2}$               &  & 2c        & $C_{2}$               &  & 4f        & $C_{2v}$              \\
					2a        & $C_{2}$               &  & 2b       & $C_{2}$               &  & 2b        & $C_{2}$               &  & 4e        & $C_{2v}$              \\
					1o        & $e$                   &  & 2a       & $C_{2}$               &  & 1a        & $e$                   &  & 4d        & $C_{2v}$              \\
					&                       &  & 1o       & $e$                   &  &           &                       &  & 2c        & $C_{2}$               \\
					&                       &  &          &                       &  &           &                       &  & 2b        & $C_{2}$               \\
					&                       &  &          &                       &  &           &                       &  & 2a        & $C_{2}$               \\
					&                       &  &          &                       &  &           &                       &  & 1o        & $e$  
					\\\hline                
				\end{tabular}
			\end{table}
			
			\begin{table}[h!]
				\begin{tabular}{ccccccccccc}
					\hline
					\multicolumn{2}{c}{$C_{4}(4)$} &  & \multicolumn{2}{c}{$S_{4}\left(\bar{4}\right)$} &  & \multicolumn{2}{c}{$C_{4h}(4/m)$} &  & \multicolumn{2}{c}{$D_{4}(422)$} \\\hline
					WP      & \quad Structure Group \quad      &  & WP               & \quad Structure Group \quad              &  & WP        & \quad Structure Group \quad       &  & WP       & \quad Structure Group \quad       \\\hline
					4b      & $C_{4}$              &  & 4b               & $S_{4}$                      &  & 8c        & $C_{4h}$              &  & 8d       & $D_{4}$               \\
					1a      & $e$                  &  & 2a               & $C_{2}$                      &  & 4b        & $C_{4}$               &  & 4c       & $C_{4}$               \\
					&                      &  & 1o               & $e$                          &  & 2a        & $C_{2}$               &  & 4b       & $C_{4}$               \\
					&                      &  &                  &                              &  & 1o        & $e$                   &  & 2a       & $C_{2}$               \\
					&                      &  &                  &                              &  &           &                       &  & 1o       & $e$       
					\\\hline           
				\end{tabular}
			\end{table}
			
			\begin{table}[h!]
				\begin{tabular}{ccccccccccc}
					\hline
					\multicolumn{2}{c}{$C_{4v}(4mm)$} &  & \multicolumn{2}{c}{$D_{2d}\left(\bar{4}2m\right)$} &  & \multicolumn{2}{c}{$D_{4h}(4/mmm)$} &  & \multicolumn{2}{c}{$C_{3}(3)$} \\\hline
					WP        & \quad Structure Group \quad       &  & WP                & \quad Structure Group \quad                &  & WP         & \quad Structure Group \quad        &  & WP      & \quad Structure Group \quad      \\\hline
					8d        & $C_{4v}$              &  & 8d                & $D_{2d}$                       &  & 16g        & $D_{4h}$               &  & 3b      & $C_{3}$              \\
					4c        & $C_{4}$               &  & 4c                & $S_{4}$                        &  & 8f         & $C_{4h}$               &  & 1a      & $e$                  \\
					4b        & $C_{4}$               &  & 4b                & $C_{4}$                        &  & 8e         & $C_{4h}$               &  &         &                      \\
					1a        & $e$                   &  & 2a                & $C_{2}$                        &  & 8d         & $D_{4}$                &  &         &                      \\
					&                       &  & 1o                & $e$                            &  & 4c         & $C_{4}$                &  &         &                      \\
					&                       &  &                   &                                &  & 4b         & $C_{4}$                &  &         &                      \\
					&                       &  &                   &                                &  & 2a         & $C_{2}$                &  &         &                      \\
					&                       &  &                   &                                &  & 1o         & $e$                    &  &         &            
					\\\hline         
				\end{tabular}
			\end{table}
			
			\begin{table}[h!]
				\begin{tabular}{ccccccccccc}
					\hline
					\multicolumn{2}{c}{$C_{3i}\left(\bar{3}\right)$} &  & \multicolumn{2}{c}{$D_{3}(32)$} &  & \multicolumn{2}{c}{$C_{3v}(3m)$} &  & \multicolumn{2}{c}{$D_{3d}\left(\bar{3}m\right)$} \\\hline
					WP               & \quad Structure Group \quad               &  & WP       & \quad Structure Group \quad      &  & WP       & \quad Structure Group \quad       &  & WP                & \quad Structure Group \quad               \\\hline
					6b               & $C_{3i}$                      &  & 6c       & $D_{3}$              &  & 6c       & $C_{3v}$              &  & 12d               & $D_{3d}$                      \\
					2a               & $C_{2}$                       &  & 3b       & $C_{3}$              &  & 3b       & $C_{3}$               &  & 6c                & $C_{3i}$                      \\
					1o               & $e$                           &  & 2a       & $C_{2}$              &  & 1a       & $e$                   &  & 6b                & $C_{6}$                       \\
					&                               &  & 1o       & $e$                  &  &          &                       &  & 2a                & $C_{2}$                       \\
					&                               &  &          &                      &  &          &                       &  & 1o                & $e$        
					\\\hline                  
				\end{tabular}
			\end{table}
			
			\begin{table}[t!]
				\begin{tabular}{ccccccccccc}
					\hline
					\multicolumn{2}{c}{$C_{6}(6)$} &  & \multicolumn{2}{c}{$C_{3h}\left(\bar{6}\right)$} &  & \multicolumn{2}{c}{$C_{6h}(6/m)$} &  & \multicolumn{2}{c}{$D_{6}(622)$} \\\hline
					WP      & \quad Structure Group \quad      &  & WP               & \quad Structure Group \quad               &  & WP        & \quad Structure Group \quad       &  & WP        & \quad Structure Group \quad      \\\hline
					6b      & $C_{6}$              &  & 6c               & $C_{3h}$                      &  & 12c       & $C_{6h}$              &  & 12d       & $D_{6}$              \\
					1a      & $e$                  &  & 3b               & $C_{3}$                       &  & 6b        & $C_{6}$               &  & 6c        & $C_{6}$              \\
					&                      &  & 2a               & $C_{2}$                       &  & 2a        & $C_{2}$               &  & 6b        & $C_{6}$              \\
					&                      &  & 1o               & $e$                           &  & 1o        & $e$                   &  & 2a        & $C_{2}$              \\
					&                      &  &                  &                               &  &           &                       &  & 1o        & $e$                 
					\\\hline
				\end{tabular}
			\end{table}
			
			\begin{table}[t]
				\begin{tabular}{ccccccccccc}
					\hline
					\multicolumn{2}{c}{$C_{6v}(6mm)$} &  & \multicolumn{2}{c}{$D_{3h}\left(\bar{6}m2\right)$} &  & \multicolumn{2}{c}{$D_{6h}(6/mmm)$} &  & \multicolumn{2}{c}{$T(23)$} \\\hline
					WP        & \quad Structure Group \quad       &  & WP                 & \quad Structure Group \quad               &  & WP         & \quad Structure Group \quad        &  & WP     & \quad Structure Group \quad    \\\hline
					12d       & $C_{6v}$              &  & 12e                & $D_{3h}$                      &  & 24g        & $D_{6h}$               &  & 12c    & $T$                \\
					6c        & $C_{6}$               &  & 6d                 & $C_{6}$                       &  & 12f        & $D_{6}$                &  & 6b     & $C_{3i}$           \\
					6b        & $C_{6}$               &  & 6c                 & $C_{3h}$                      &  & 12e        & $C_{6h}$               &  & 4a     & $S_{4}$            \\
					1a        & $e$                   &  & 3b                 & $C_{3}$                       &  & 12d        & $C_{6h}$               &  & 1o     & $e$                \\
					&                       &  & 2a                 & $C_{2}$                       &  & 6c         & $C_{6}$                &  &        &                    \\
					&                       &  & 1o                 & $e$                           &  & 6b         & $C_{6}$                &  &        &                    \\
					&                       &  &                    &                               &  & 2a         & $C_{2}$                &  &        &                    \\
					&                       &  &                    &                               &  & 1o         & $e$                    &  &        &          
					\\\hline         
				\end{tabular}
			\end{table}
			
			\begin{table}[t]
				\begin{tabular}{ccccccccccc}
					\hline
					\multicolumn{2}{c}{$T_{h}\left(m\bar{3}\right)$} &  & \multicolumn{2}{c}{$O(432)$} &  & \multicolumn{2}{c}{$T_{d}\left(\bar{4}3m\right)$} &  & \multicolumn{2}{c}{$O_{h}\left(m\bar{3}m\right)$} \\\hline
					WP                & \quad Structure Group \quad              &  & WP      & \quad Structure Group \quad    &  & WP                & \quad Structure Group \quad               &  & WP                & \quad Structure Group \quad               \\\hline
					24d               & $T_{h}$                      &  & 24d     & $O$                &  & 24d               & $T_{d}$                       &  & 48f               & $O_{h}$                       \\
					12c               & $T$                          &  & 12c     & $T$                &  & 12c               & $T$                           &  & 24e               & $O$                           \\
					8b                & $C_{4h}$                     &  & 8b      & $C_{4h}$           &  & 6b                & $C_{3i}$                      &  & 24d               & $O$                           \\
					6a                & $C_{3i}$                     &  & 6a      & $C_{3i}$           &  & 4a                & $S_{4}$                       &  & 12c               & $T$                           \\
					1o                & $e$                          &  & 1o      & $e$                &  & 1o                & $e$                           &  & 8b                & $C_{4h}$                      \\
					&                              &  &         &                    &  &                   &                               &  & 6a                & $C_{3i}$                      \\
					&                              &  &         &                    &  &                   &                               &  & 1o                & $e$         
					\\\hline                 
				\end{tabular}
			\end{table}
			
		\end{widetext}
	\end{appendix}	
	
\end{document}